\documentclass[10pt]{article}

\PassOptionsToPackage{table}{xcolor}

\usepackage{scai}

\definecolor{promptblue}{HTML}{2B5B9E}
\definecolor{prompttea}{HTML}{1F7A6F}
\definecolor{promptplum}{HTML}{8A5CD6}
\definecolor{promptamber}{HTML}{B45A1C}
\definecolor{promptolive}{HTML}{556B2F}
\definecolor{citepink}{HTML}{E63E91}

\hypersetup{
  colorlinks=true,
  citecolor=promptblue,
  linkcolor=promptplum,
  urlcolor=citepink,
  pdftitle={From Visual Synthesis to Interactive Worlds: Toward Production-Ready 3D Asset Generation},
  pdfauthor={Jiafeng Wu, Zhuofan Lou, Jian Liu, Dazhao Du, Chunchao Guo, Song Guo},
  pdfsubject={Production-ready 3D asset generation for interactive worlds and asset pipelines},
  pdfkeywords={production-ready 3D generation, asset production pipeline, interactive worlds, mesh processing, retopology, UV unwrapping, PBR materials, rigging and skinning, 3D scene assembly}
}

\usepackage[authoryear, sort&compress, round]{natbib}

\usepackage{booktabs}
\usepackage{multirow}
\usepackage{tabularx}
\newcolumntype{C}{>{\centering\arraybackslash}X}
\usepackage{threeparttable}
\usepackage{pifont}
\usepackage{tikz}
\usetikzlibrary{calc,positioning}
\usepackage{nicefrac}
\usepackage{placeins}
\usepackage{wrapfig}
\usepackage{listings}
\usepackage{float}

\AtBeginDocument{%
  }
\providecommand{\Description}[1]{}

\makeatletter
\g@addto@macro\normalsize{%
  \setlength{\abovedisplayskip}{3pt plus 1pt minus 1pt}%
  \setlength{\abovedisplayshortskip}{0pt plus 1pt}%
  \setlength{\belowdisplayskip}{3pt plus 1pt minus 1pt}%
  \setlength{\belowdisplayshortskip}{2pt plus 1pt minus 1pt}%
}
\g@addto@macro\small{%
  \setlength{\abovedisplayskip}{2pt plus 1pt minus 1pt}%
  \setlength{\abovedisplayshortskip}{0pt plus 1pt}%
  \setlength{\belowdisplayskip}{2pt plus 1pt minus 1pt}%
  \setlength{\belowdisplayshortskip}{1pt plus 1pt minus 1pt}%
}
\makeatother
\normalsize

\definecolor{grpSdsHead}{HTML}{CD9B9B}
\definecolor{grpSdsRow}{HTML}{FAF3F3}
\definecolor{grpMvHead}{HTML}{C199B0}
\definecolor{grpMvRow}{HTML}{F9F3F6}
\definecolor{grpGanHead}{HTML}{89ADD2}
\definecolor{grpGanRow}{HTML}{F2F6FA}
\definecolor{grpVaeHead}{HTML}{8BBFA5}
\definecolor{grpVaeRow}{HTML}{F2F8F5}
\definecolor{grpDdmHead}{HTML}{CCB583}
\definecolor{grpDdmRow}{HTML}{FAF7F0}
\definecolor{grpFfHead}{HTML}{A99CC5}
\definecolor{grpFfRow}{HTML}{F5F3F9}
\definecolor{grpLgmHead}{HTML}{80BCC4}
\definecolor{grpLgmRow}{HTML}{F1F7F8}
\definecolor{colHead}{HTML}{5B7282}

\definecolor{grpParamHead}{HTML}{C99BA0}
\definecolor{grpParamRow}{HTML}{FAF2F3}
\definecolor{grpImplHead}{HTML}{A89DC5}
\definecolor{grpImplRow}{HTML}{F4F2F9}
\definecolor{grpGenHead}{HTML}{8EBB9A}
\definecolor{grpGenRow}{HTML}{F1F8F3}
\definecolor{grpFfbHead}{HTML}{82B4D4}
\definecolor{grpFfbRow}{HTML}{F0F6FA}
\definecolor{grpMorphHead}{HTML}{C5AD83}
\definecolor{grpMorphRow}{HTML}{FAF7F0}
\definecolor{grpGsHead}{HTML}{7DB8AE}
\definecolor{grpGsRow}{HTML}{F0F7F6}
\definecolor{grpFfrecHead}{HTML}{CDA382}
\definecolor{grpFfrecRow}{HTML}{FBF6F0}
\definecolor{grpAnimHead}{HTML}{B694B0}
\definecolor{grpAnimRow}{HTML}{F8F2F7}

\definecolor{grpLayHead}{HTML}{8AACC5}
\definecolor{grpLayRow}{HTML}{F1F5FA}
\definecolor{grpGndHead}{HTML}{8BBF9A}
\definecolor{grpGndRow}{HTML}{F1F8F3}
\definecolor{grpWldHead}{HTML}{A093C0}
\definecolor{grpWldRow}{HTML}{F4F2F8}

\newcommand{\emailicon}{%
  \tikz[baseline=-0.28ex, x=1em, y=1em]{
    \fill[promptblue!85!black, rounded corners=0.12em]
      (0,0) rectangle (1.12,0.82);
    \fill[white, rounded corners=0.05em]
      (0.16,0.18) rectangle (0.96,0.64);
    \draw[promptblue!85!black, line width=0.28pt]
      (0.17,0.62) -- (0.56,0.36) -- (0.95,0.62);
    \draw[promptblue!70!black, line width=0.24pt]
      (0.17,0.19) -- (0.46,0.43);
    \draw[promptblue!70!black, line width=0.24pt]
      (0.95,0.19) -- (0.66,0.43);
  }%
}
\newcommand{\contribbadge}[2]{%
  \tcbox[
    on line,
    enhanced,
    colback=promptblue!5!white,
    colframe=promptblue!55!black,
    boxrule=0.35pt,
    arc=2pt,
    left=4pt,
    right=4pt,
    top=1.5pt,
    bottom=1.5pt,
    boxsep=0pt,
  ]{\itshape\textsuperscript{#1}#2}%
}

\renewcommand{\authorfont}{\normalsize\rmfamily}

\makeatletter
\renewcommand{\scaiauthor}[2]{%
  \stepcounter{scai@numauthors}%
  \ifnum\value{scai@numauthors}=1
    \gdef\@scai@authorlist{\mbox{\textbf{#2}\textsuperscript{\@scai@rendertags{#1}}}}%
  \else
    \g@addto@macro\@scai@authorlist{, \mbox{\textbf{#2}\textsuperscript{\@scai@rendertags{#1}}}}%
  \fi
}

\renewcommand{\@scai@renderaffils}{%
  \begingroup
  \normalsize
  \makebox[\textwidth][c]{\textsuperscript{1}\csname @scai@affil@1\endcsname}\par
  \makebox[\textwidth][c]{\textsuperscript{2}\csname @scai@affil@2\endcsname}\par
  \makebox[\textwidth][c]{%
    \textsuperscript{3}\csname @scai@affil@3\endcsname
    \quad
    \textsuperscript{4}\csname @scai@affil@4\endcsname
  }\par
  \endgroup
}
\makeatother

\makeatletter
\newcommand{\authorbreak}{%
  \g@addto@macro\@scai@authorlist{\\}%
  \let\@scai@authorsaved\scaiauthor
  \def\scaiauthor##1##2{%
    \stepcounter{scai@numauthors}%
    \g@addto@macro\@scai@authorlist{\mbox{\textbf{##2}\textsuperscript{\@scai@rendertags{##1}}}}%
    \let\scaiauthor\@scai@authorsaved
  }%
}
\makeatother

\makeatletter
\fancypagestyle{scaifirstpage}{%
  \fancyhf{}%
  \fancyhead[L]{%
    \includegraphics[height=31pt, trim=28.44pt 740.88pt 519.84pt 28.44pt, clip]{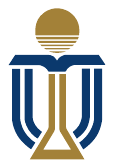}%
    \hspace{0.9em}%
    \includegraphics[height=31pt, trim=28.44pt 762.84pt 500.4pt 28.8pt, clip]{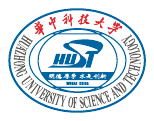}%
    \hspace{0.9em}%
    \includegraphics[height=31pt, trim=28.44pt 151.2pt 27.72pt 151.2pt, clip]{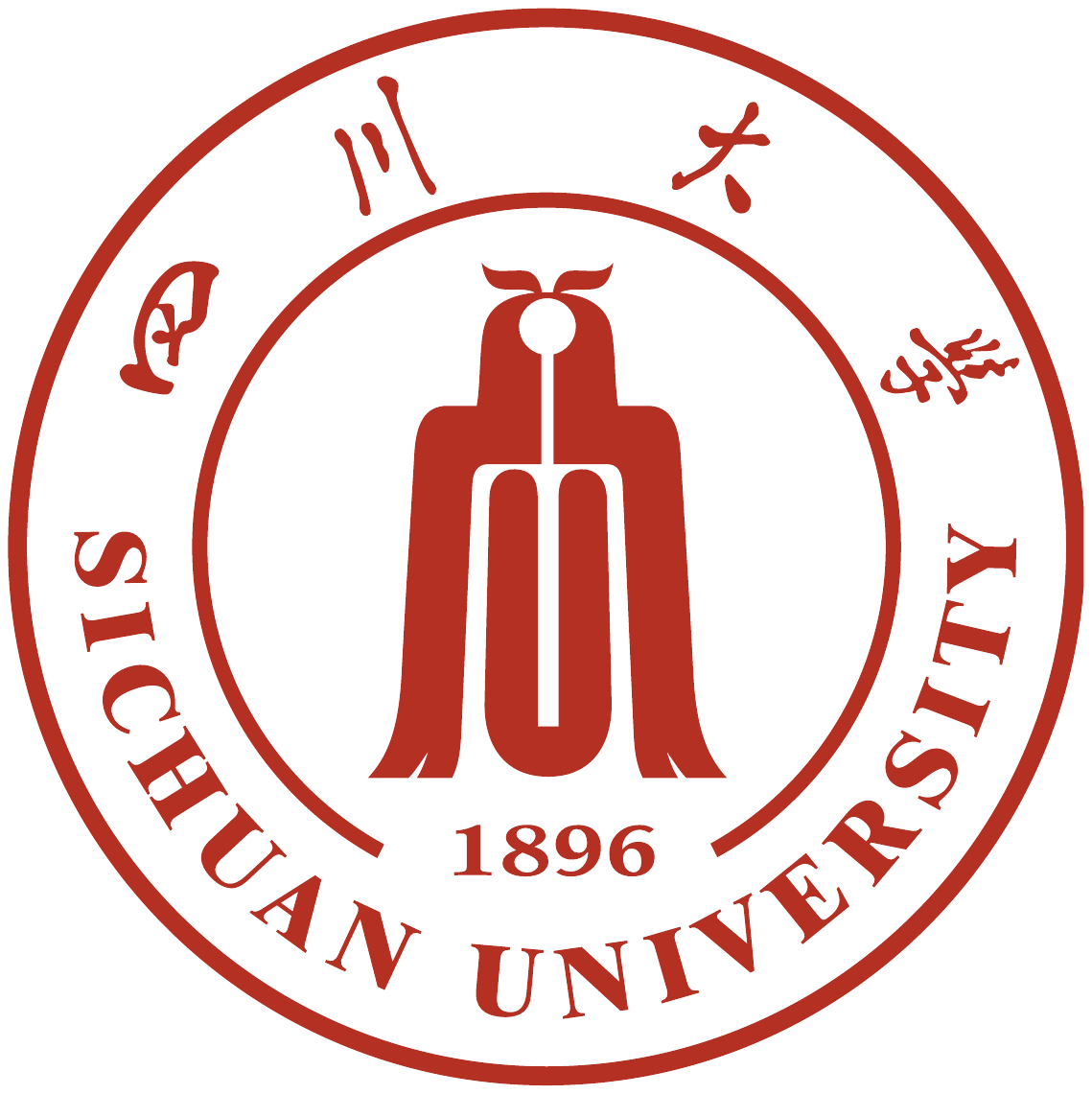}%
  }%
  \fancyfoot[R]{\footerfont\thepage}%
  \renewcommand{\headrule}{\color{gray}\@scai@DefaultHeadRule}%
}

\renewcommand{\maketitle}{%
  \bgroup
  \setlength{\parindent}{0pt}%
  \setlength{\parskip}{0pt}%
  \vspace*{3pt}%
  {\centering\titlefont\@title\par}%
  \vspace{14pt}%
  {\centering\authorfont\@scai@authorlist\par}%
  \vspace{8pt}%
  {\centering\@scai@renderaffils\par}%
  \vspace{5pt}%
  {\centering\small
    \contribbadge{$\ast$}{Equal contribution.}\quad
    \contribbadge{$\S$}{Work done during internship at HKUST.}\quad
    \contribbadge{$\ddagger$}{Project lead.}\quad
    \contribbadge{$\dagger$}{Corresponding author.}\par}%
  \if@scai@haskeywords
    \vspace{6pt}%
    {\centering\keywordsfont{\sffamily\textbf{Keywords:}} \@scai@keywords\par}%
  \fi
  \vspace{10pt}%
  \@scai@renderabstract
  \egroup
  \thispagestyle{scaifirstpage}%
}
\makeatother

\title{From Visual Synthesis to Interactive Worlds:\\Toward Production-Ready 3D Asset Generation}

\contribnote{equal}{$\ast$}{Equal contribution.}
\contribnote{intern}{$\S$}{Work done during internship at HKUST.}
\contribnote{lead}{$\ddagger$}{Project lead.}
\contribnote{corr}{$\dagger$}{Corresponding author.}
\makeatletter
\renewcommand{\@scai@rendercontribfooter}{}
\makeatother

\scaiauthor{1,2,equal,intern}{Jiafeng Wu}
\scaiauthor{1,3,equal,intern}{Zhuofan Lou}
\scaiauthor{1,lead}{Jian Liu}
\scaiauthor{1}{Dazhao Du}
\scaiauthor{4}{Chunchao Guo}
\scaiauthor{1,corr}{Song Guo}

\affiliation{1}{The Hong Kong University of Science and Technology}
\affiliation{2}{Huazhong University of Science and Technology}
\affiliation{3}{Sichuan University}
\affiliation{4}{Independent Researcher}

\metadata[\emailicon\,Email]{\href{mailto:dyadavev@gmail.com}{dyadavev@gmail.com}, \href{mailto:zhuofanlou2005@gmail.com}{zhuofanlou2005@gmail.com}, \href{mailto:jliuhr@connect.ust.hk}{jliuhr@connect.ust.hk}, \href{mailto:songguo@cse.ust.hk}{songguo@cse.ust.hk}}
\metadata[\raisebox{-0.35ex}{\includegraphics[height=1.25em]{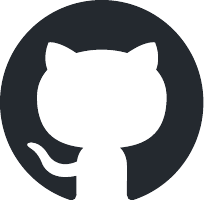}}\,Project page]{\url{https://christinebobby.github.io/production-ready-3d-survey/}}

\scaikeywords{production-ready 3D generation, asset production pipeline, interactive worlds, mesh processing, retopology, UV unwrapping, PBR materials, texture synthesis, rigging and skinning, 3D scene assembly, world models}

\begin{abstract}
Three-dimensional content generation has progressed from producing isolated, visually plausible shapes to constructing structured assets that can be deployed in real-time interactive environments. This trajectory is driven by converging demands from game development, embodied AI, world simulation, digital twins, and spatial computing, all of which require 3D content that goes beyond surface appearance to satisfy engine-level constraints on topology, UV parameterization, physically based materials, skeletal rigging, and physics-aware scene layout. Despite rapid advances in generative modeling, a persistent gap separates the outputs of current methods from the production-ready standard expected by interactive applications. This survey addresses that gap by organizing the literature around the asset production pipeline rather than algorithmic families. Along the horizontal axis we distinguish three asset tiers, namely general objects, characters, and scenes, while the vertical axis traces each tier through the full production lifecycle from data foundations and geometry synthesis through topology optimization, UV unwrapping, PBR appearance, rigging, and scene assembly. Through this two-dimensional taxonomy we assess not only what current methods can generate but whether their outputs are directly usable in downstream engines and simulation platforms. We further consolidate evaluation metrics and protocols that span geometric fidelity, appearance quality, asset usability, and scene-level physical plausibility. The survey concludes by identifying open challenges in data quality, generation controllability, end-to-end assetization, and physically grounded generation, and by situating production-ready 3D content as foundational infrastructure for emerging interactive world models and embodied intelligent systems.
\end{abstract}

\begin{document}

\maketitle

\section{Introduction}
Three-dimensional content generation has advanced rapidly, with score distillation, latent 3D diffusion, and feed-forward reconstruction~\cite{poole2023dreamfusion,lin2023magic3d,wang2023prolificdreamer,zhao2023michelangelo,zhang2024clay,zhao2025hunyuan3d,hong2024lrm,zhang2024gs,li2025triposg} raising the visual quality, diversity, and speed of 3D asset creation to unprecedented levels. Yet the field now confronts a question that visual fidelity alone cannot answer: can these generated assets function as first-class components of interactive systems? Applications in world simulation, embodied intelligence, digital twins, and spatial computing do not merely consume 3D geometry as static imagery; they require structured, production-ready content that can be rendered in real time, physically simulated, animated, and edited within interactive environments.

The breadth of this demand is becoming increasingly clear. Video-based world models~\cite{bruce2024genie,valevski2025gamengen,nvidia2025cosmos} achieve impressive visual realism but operate without explicit 3D structure: their outputs are not structured to support decomposition into editable objects or explicit physical guarantees. Structured approaches---procedural generation~\cite{raistrick2023infinite,deitke2022procthor} and LLM-driven composition~\cite{yang2024holodeck,sun20253d,hu2024scenecraft}---provide editability and physical grounding but remain constrained by the availability of production-quality assets, the same bottleneck limiting embodied AI platforms~\cite{savva2019habitat,makoviychuk2021isaacgym,deitke2022procthor,raistrick2024infinigenindoors}. Across these settings, the underlying requirement converges on a common set of production-level attributes: clean mesh topology, proper UV parameterization, physically based materials separated from baked lighting, skeletal rigs for articulated animation, and collision geometry for physical interaction.

A persistent gap, however, separates the outputs of current generative methods from these requirements. Score distillation approaches~\cite{poole2023dreamfusion,lin2023magic3d,wang2023prolificdreamer} produce geometry through iterative optimization that frequently yields noisy surfaces with lighting entangled in the albedo channel~\cite{zeng2024paint3d}. Feed-forward reconstruction models~\cite{hong2024lrm,zhang2024gs,TripoSR2024} generate implicit or semi-implicit representations that require non-trivial extraction and cleanup before engine import. Autoregressive mesh generators~\cite{siddiqui2024meshgpt,chen2024meshanything,zhao2025deepmesh} have begun to address topology directly, yet their outputs typically lack UV maps, PBR material decomposition, and skeletal rigging. Recent integrated systems~\cite{lei2025hunyuan3d,xiang2025structured} combine multiple pipeline stages but remain limited in asset scope or still require manual intervention at key steps. In practice, converting a generated 3D object into a production-ready asset involves retopology, UV layout, material authoring, and rigging, a sequence of labor-intensive steps that collectively constitute what we term the assetization bottleneck.

Game engines such as Unreal Engine and Unity impose the strictest requirements on 3D assets, offering the clearest definition of what \emph{production-ready} concretely means~\cite{lei2025hunyuan3d,zhao2025hunyuan3d}. An asset reaches production readiness only when it can be deployed directly in an engine without manual repair, which requires: \textbf{(1)~Manifold mesh topology}---watertight surfaces with structured edge loops suitable for deformation and LOD management~\cite{liu2025quadgpt,siddiqui2024meshgpt,lei2025hunyuan3d}; \textbf{(2)~Non-overlapping UV parameterization}---a distortion-controlled atlas where every texel maps unambiguously to a surface point~\cite{li2025auto}; \textbf{(3)~Disentangled PBR materials}---illumination-independent albedo, roughness, metallic, and normal maps that relight correctly under arbitrary engine lighting~\cite{zeng2024paint3d,yu2024texgen,he2025materialmvp,zhao2025hunyuan3d}; \textbf{(4)~Skeletal rig and skinning weights} for animatable assets, compatible with the engine's animation graph~\cite{xu2020rignet,mosella2022skinningnet}; \textbf{(5)~Post-generation editability}---geometry, materials, and skeleton must remain independently accessible~\cite{xiang2025trellis2,lei2025hunyuan3d}; and \textbf{(6)~Physics metadata}---collision meshes, mass, and friction parameters for rigid-body simulation~\cite{yang2024physcene,nie2022pose2room,yi2023mime}. These attributes specify structural correctness for \emph{function} inside interactive systems, and expose why current benchmarks---which omit topology, UV, rigging, and engine import metrics~\cite{tang2025recent,fukaya2025intelligent}---systematically overestimate deployment readiness. This operational definition serves as the evaluative lens throughout the survey.
Existing surveys on 3D generative AI typically adopt a taxonomy centered on algorithmic families, reviewing the progression from GANs and VAEs through diffusion models to autoregressive architectures~\cite{tang2025recent,foo2025ai}. Others focus specifically on scene generation~\cite{wen20253d} or provide high-level analyses of AI-generated game content~\cite{fukaya2025intelligent}. While these works offer valuable references for understanding the underlying algorithms, their model-centric organization does not address the complete lifecycle of an asset from synthesis through engine deployment. They provide little guidance on where the critical deployment bottlenecks lie, whether in topology extraction, material disentanglement, automatic rigging, or scene-level physical plausibility, making it difficult to target the stages that matter most for practical impact.

This survey addresses that gap by organizing the literature around the production pipeline itself. We construct a two-dimensional taxonomy in which the horizontal axis distinguishes three asset tiers (general objects and props, characters and avatars, and scenes and environments) and the vertical axis traces each tier through the full production lifecycle: data foundations and representations, geometry synthesis, topology optimization and UV unwrapping, PBR appearance generation, skeletal rigging, and scene assembly. Through this cross-sectional structure, we assess not only what current methods can generate but whether their outputs are directly usable in interactive engines and simulation platforms. We further situate production-ready 3D content as foundational infrastructure for the emerging convergence of structured world models, embodied intelligent systems, and spatial computing.

\begin{figure*}[t]
    \centering
    \includegraphics[width=0.88\textwidth]{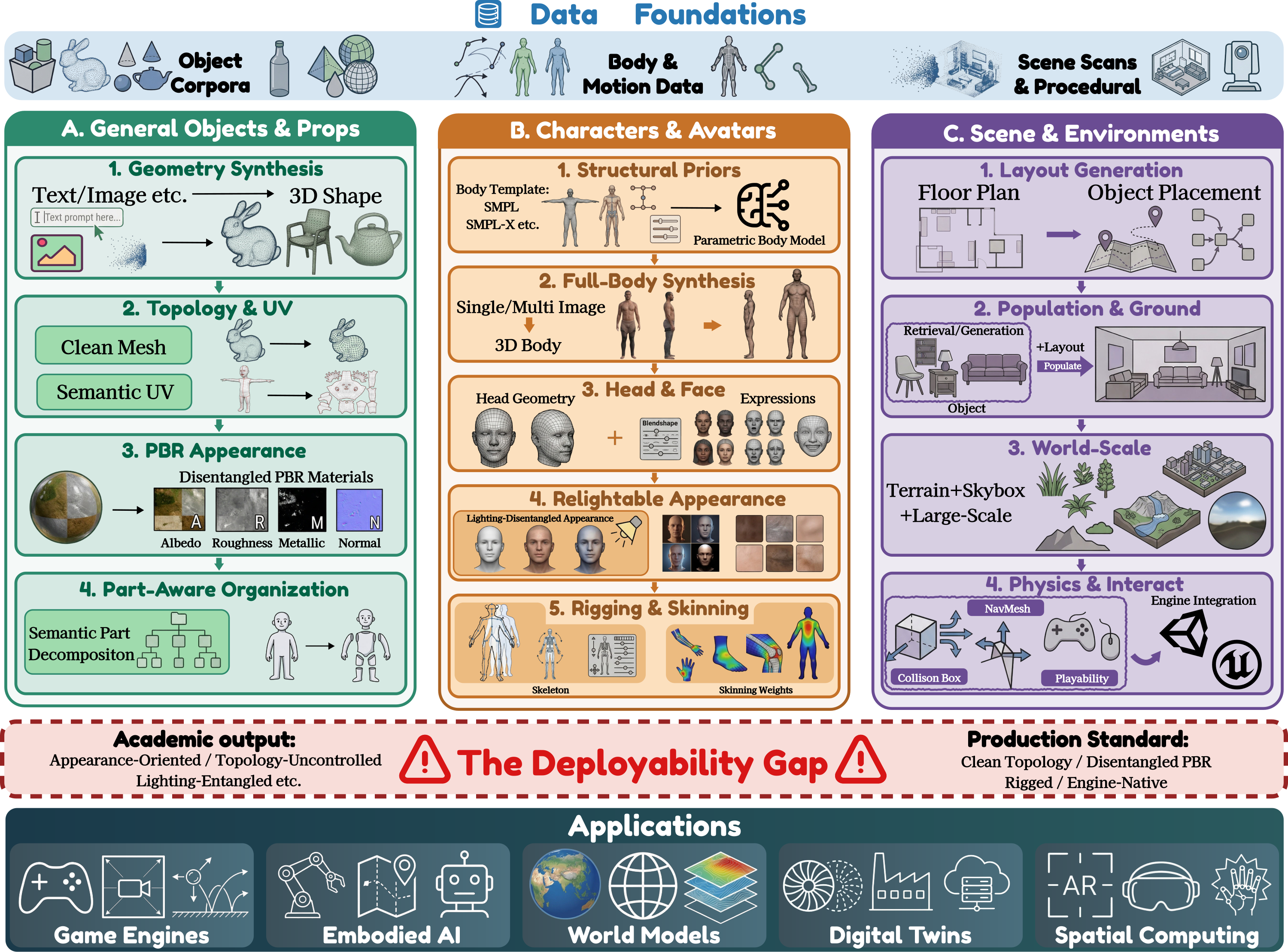}
    \caption{Pipeline-centric taxonomy bridging 3D synthesis and production-ready assets. Vertical axis: workflow stages; horizontal axis: asset categories (objects, characters, scenes). The bridge motif highlights the gap between appearance-first research and production requirements.}
    \Description{A two-dimensional taxonomy diagram titled Bridging the Gap: From Isolated 3D Appearance to Production-Ready Usability. The left axis lists four industrial production workflow stages: data foundations and representations, geometry synthesis, topology and UV unwrapping, and PBR appearance synthesis. The top axis labels three asset categories: general objects and props, characters and avatars, and scenes and environments. A central bridge motif connects academic focus on isolated geometry to industrial requirements. Detailed visual icons illustrate category-specific production steps including quad-mesh topology, PBR texture maps, body models, skeletal rigs, skinning weights, layout generation, collision meshes, and LLM-driven scene assembly. A bottom panel summarizes core survey contributions.}
    \label{fig:taxonomy}
    \vspace{-10pt}
\end{figure*}

The contributions of this survey are fourfold. First, we reorganize 3D generation research around the game asset production lifecycle, mapping academic tasks to industrial workflow stages. Second, we provide focused reviews of critical post-generation stages---topology generation~\cite{liu2025quadgpt,siddiqui2024meshgpt}, UV parameterization~\cite{li2025auto}, PBR material prediction~\cite{zeng2024paint3d,yu2024texgen,he2025materialmvp}, and automatic rigging~\cite{xu2020rignet,mosella2022skinningnet}---that currently hinder deployment. Third, we trace the evolution of 3D datasets from ShapeNet~\cite{chang2015shapenet} through Objaverse~\cite{deitke2023objaverse,deitke2023objaversexl} to procedural environments~\cite{deitke2022procthor,raistrick2023infinite}, and examine how data representations~\cite{loper2015smpl,fu20213d,paschalidou2021atiss} determine controllability and editability. Fourth, we identify open challenges in physics-aware generation~\cite{yang2024physcene,chen2025physgen3d}, LLM-driven scene automation~\cite{sun20253d,hu2024scenecraft,yang2024holodeck,deng2024citycraft,tang2025unrealllm}, and end-to-end assetization, and discuss the role of production-ready 3D content for world models and embodied intelligence.

To maintain a focused narrative, we explicitly define what is out of scope. We do not cover 2D generative techniques used solely for conceptual art or UI elements, unless they serve as condition inputs for 3D lifting. Purely offline rendering algorithms targeted at VFX production, which do not conform to the constraints of real-time engines, are excluded. Fundamental improvements in generic diffusion sampling algorithms or the underlying mathematics of implicit fields are omitted unless they directly resolve specific bottlenecks in the production pipeline of 3D assets.

\textbf{Paper Organization.} \hyperref[sec:background]{Section~\ref*{sec:background}} reviews the foundational 3D representations and generative paradigms that underpin the methods discussed in later sections. \hyperref[sec:data_foundations]{Section~\ref*{sec:data_foundations}} surveys data foundations and benchmarks across all asset categories, identifying structural gaps relevant to production-ready generation. \hyperref[sec:general_objects]{Section~\ref*{sec:general_objects}} addresses General Objects and Props, tracing the production pipeline from geometry synthesis through topology optimization, UV unwrapping, and PBR material generation. \hyperref[sec:characters_and_avatars]{Section~\ref*{sec:characters_and_avatars}} turns to Characters and Avatars, covering parametric body priors, animatable avatar synthesis, and automatic skeletal rigging. \hyperref[sec:scenes_environments]{Section~\ref*{sec:scenes_environments}} examines Scenes and Environments, with attention to structural layout, physical plausibility, and interactive world construction. \hyperref[sec:evaluation]{Section~\ref*{sec:evaluation}} consolidates evaluation metrics and protocols for geometric fidelity, appearance quality, asset usability, and scene-level assessment. \hyperref[sec:open_challenges]{Section~\ref*{sec:open_challenges}} discusses open challenges and future directions, including data quality, controllability, end-to-end assetization, physically grounded generation, and world models. \hyperref[sec:conclusion]{Section~\ref*{sec:conclusion}} concludes the survey.

\begin{figure*}[t]
    \centering
    \includegraphics[width=0.99\textwidth]{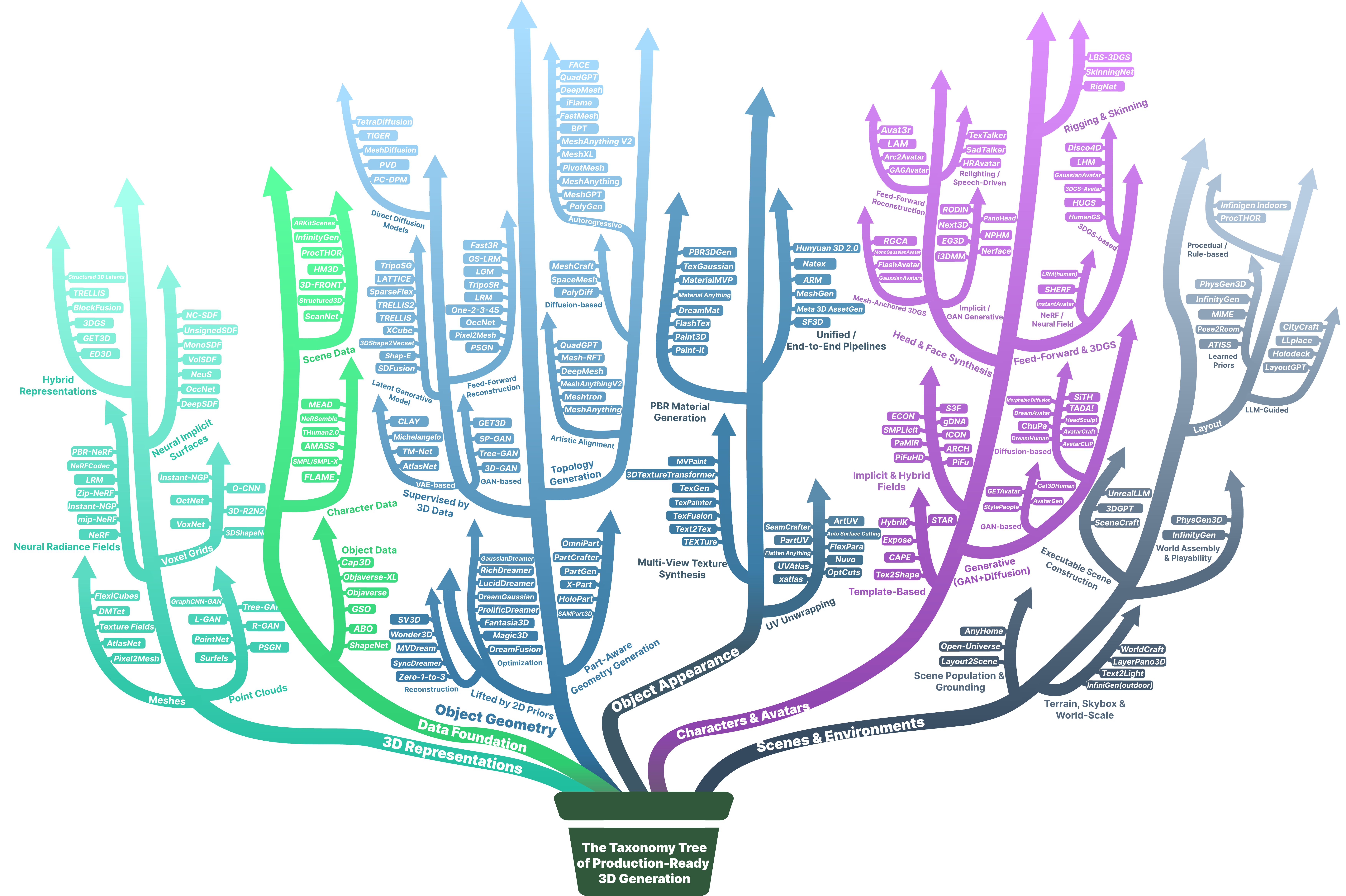}
    \caption{Method taxonomy tree spanning five survey branches: 3D representations, data foundations, object geometry, object appearance, and characters \& avatars / scenes \& environments. Leaf nodes denote representative methods.}
    \Description{A taxonomy tree diagram showing production-ready 3D generation methods organized into five main branches growing upward from a common trunk labeled ``The Taxonomy Tree of Production-Ready 3D Generation.'' From left to right: 3D Representations (neural radiance fields, implicit surfaces, hybrid representations, voxel grids, meshes, point clouds), Data Foundation (scene data, character data, object data), Object Geometry (direct diffusion models, feed-forward reconstruction, GAN/VAE-based, optimization), Object Appearance (PBR material generation, topology generation, UV unwrapping, multi-view texture synthesis), Characters \& Avatars (head \& face synthesis, feed-forward \& 3DGS, rigging \& skinning), and Scenes \& Environments (layout \& grounding, terrain \& world-scale, executable scene construction, LLM-guided).}
    \label{fig:taxonomy_tree}
\vspace{-10pt}
\end{figure*}

\section{Background: 3D Representations and Generative Paradigms}
\label{sec:background}
Two closely related questions lie at the core of modern 3D asset generation: how three-dimensional content is represented, and how such representations are learned or produced from data. The first concerns the underlying data structure; the second concerns the generative paradigm. This distinction matters because representation and paradigm jointly determine the fidelity, controllability, and deployability of the resulting assets. This section introduces the principal 3D representations and generative paradigms that underpin the discussions that follow.

\subsection{3D Representations}

\paragraph{Point Clouds.} A point cloud $\mathcal{P} = \{p_i \in \mathbb{R}^{3+c}\}$ stores spatial coordinates and optional attributes (normals, colors, reflectance)~\cite{guo2020deep, qi2017pointnet}. This format aligns with sensor outputs but lacks connectivity, manifoldness, and watertightness~\cite{guo2020deep}, making it unsuitable for UV parameterization, rigging, or collision mesh construction.

\paragraph{Voxel Grids.} A voxel grid $\mathcal{V} = \{v_{i,j,k} \in \mathbb{R}^{c}\}$ discretizes 3D space into a regular $H \times W \times D$ lattice storing occupancy, density, or feature vectors~\cite{guo2020deep, wu2015modelnet, Maturana2015VoxNet}. Regularity enables direct 3D convolutions, but cubic $O(N^3)$ scaling limits resolution~\cite{guo2020deep, tatarchenko2017octree}. Sparse and hierarchical variants---octrees~\cite{riegler2017octnet}, structured latent hierarchies~\cite{muller2022instant, ren2024xcube, xiang2025structured, xiang2025trellis2}---redistribute computation to occupied regions, though the trade-off between surface detail and efficiency remains a common concern.

\paragraph{Meshes.} A mesh $\mathcal{M} = (\mathcal{V}, \mathcal{E}, \mathcal{F})$ consists of vertices, edges, and faces that explicitly encode surface connectivity~\cite{ahmed2018survey}. This structure supports local editing, subdivision, deformation, UV parameterization, rigging, and animation within standard digital content creation and game engine toolchains, making meshes the most deployment-oriented primary representation. The trade-off is modeling complexity: meshes are irregular and non-Euclidean, which limits their compatibility with standard grid-based neural operators~\cite{ahmed2018survey, monti2017geometric}, and generating valid meshes requires consistent treatment of both vertex geometry and surface connectivity. These challenges are particularly pronounced when topology varies across categories or when thin structures and high-curvature regions must be preserved, yet meshes remain the canonical representation for production-ready assets.

\paragraph{Neural Radiance Fields.} A NeRF is a continuous neural field $F_{\theta}:\mathbb{R}^{3}\times\mathbb{S}^{2}\rightarrow\mathbb{R}^{+}\times\mathbb{R}^{3}$ mapping spatial location and viewing direction to density and color~\cite{mildenhall2021nerf}. NeRF enables high-fidelity novel-view synthesis~\cite{barron2022mip, barron2023zip} but encodes geometry only implicitly, providing no explicit surfaces for collision, UV construction, or animation. Volumetric ray marching is costly relative to rasterization~\cite{muller2022instant}, and geometry, material, and illumination are typically entangled. NeRF remains an appearance-first representation with limited suitability for production deployment.

\paragraph{Neural Implicit Surfaces.} A neural implicit surface is defined by a scalar field $f_{\theta}:\mathbb{R}^{3}\rightarrow\mathbb{R}$ whose zero level set $\mathcal{S}=\{\mathbf{x}\mid f_{\theta}(\mathbf{x})=0\}$ defines the surface~\cite{park2019deepsdf, mescheder2019occupancy}. SDF formulations dominate due to greater geometric expressiveness~\cite{park2019deepsdf, wang2021neus, yariv2021volume}. The principal advantage is topological flexibility: continuous fields enable resolution-independent, smooth boundary recovery without mesh templates~\cite{mescheder2019occupancy}. However, mesh extraction via isosurface methods~\cite{LorensenC87} is required before downstream asset processing.

\paragraph{Hybrid Representations.}
Hybrid representations combine complementary primitives, $\mathcal{H} = (\mathcal{R}_1,\mathcal{R}_2,\phi)$, where $\phi$ coordinates geometry or appearance between components. Three patterns dominate: (1)~explicit scaffolds with neural appearance, where a mesh provides animation-ready structure while neural textures or Gaussians encode view-dependent detail~\cite{shao2024splattingavatar, wang2025mega}; (2)~implicit fields with differentiable extraction, where DMTet~\cite{shen2021deep} and FlexiCubes~\cite{shen2023flexicubes} yield meshes while retaining topological flexibility~\cite{gao2022get3d}; and (3)~3D Gaussian Splatting~\cite{kerbl20233d}, which renders anisotropic Gaussian primitives via differentiable tile-based rasterization for real-time speeds. Structured latent methods organize information into tri-planes or sparse grids~\cite{chan2022efficient, xiang2025structured}, providing scalable generation backbones.

\begin{figure*}[t]
    \centering
    \includegraphics[width=0.90\textwidth]{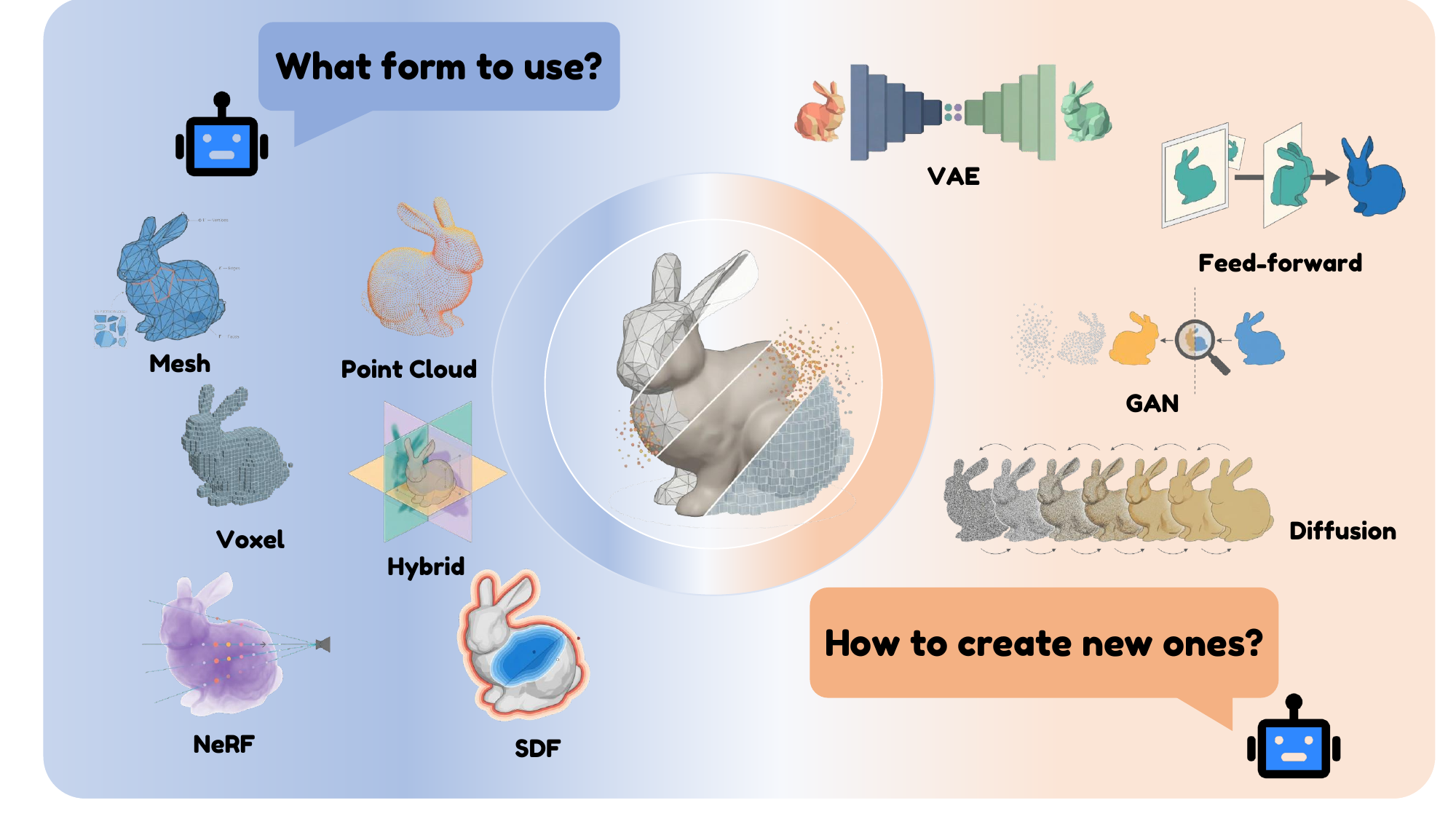}
    \caption{Taxonomy of 3D shape representations and corresponding generative modeling paradigms.}
    \label{fig:representation_taxonomy}
\end{figure*}

\subsection{Generative Paradigms}
While 3D representations determine how geometry or appearance is encoded, generative paradigms determine how such representations are learned, sampled, or reconstructed from data. Paradigms are introduced here at the level of learning objective and generation mechanism, two factors that largely govern controllability, fidelity, and scalability. Specific methods appear in later sections.

\paragraph{Variational Autoencoding Paradigms.}
Variational autoencoding paradigms~\cite{kingma2013auto} introduce a latent variable $z$ and optimize the evidence lower bound,
\[
\mathcal{L}_{\mathrm{VAE}}(x)
=
\mathbb{E}_{q_{\phi}(z|x)}[\log p_{\theta}(x|z)]
-
\mathrm{KL}\!\left(q_{\phi}(z|x)\,\|\,p(z)\right),
\]
where $q_{\phi}(z|x)$ is the encoder and $p_{\theta}(x|z)$ the decoder. The KL term regularizes the posterior, inducing structured latent embeddings. In 3D generation, this yields compact continuous codes decodable into point sets, voxels, meshes, or implicit fields.

\paragraph{Adversarial Paradigms.}
Adversarial paradigms~\cite{goodfellow2014generative} formulate generation as a minimax game between a generator $G$ and a discriminator $D$:
\[
\min_{G}\max_{D}
\;
\mathbb{E}_{x\sim p_{\mathrm{data}}}[\log D(x)]
+
\mathbb{E}_{z\sim p(z)}[\log(1-D(G(z)))].
\]
The generator maps latent $z$ to $G(z)$; the discriminator distinguishes real from generated samples. This implicit distribution matching drives strong perceptual realism but is prone to training instability and mode collapse.

\paragraph{Diffusion-Based Paradigms.}
Diffusion-based paradigms~\cite{ho2020denoising} define generation through a gradual forward corruption process and a learned reverse denoising process. Starting from a clean sample $x_0$, Gaussian noise is progressively added across $T$ steps. Generation is learned by training a denoising network $\epsilon_{\theta}$ to predict the injected noise:
\[
\mathcal{L}_{\mathrm{DM}}
=
\mathbb{E}_{x_0,\epsilon,t}
\left[
\left\|
\epsilon-\epsilon_{\theta}(x_t,t,c)
\right\|_2^2
\right],
\]
where $c$ is an optional conditioning signal. Iterative denoising provides stable training and flexible conditioning, making diffusion effective for conditional 3D generation.

\paragraph{Feed-Forward Paradigms.}
Feed-forward paradigms~\cite{fan2017point, wang2018pixel2mesh, hong2024lrm, zhang2024gs} cast 3D generation as a direct mapping $\hat{y} = F_{\theta}(c)$ from input condition $c$ to predicted output, trained with a reconstruction objective:
\[
\mathcal{L}_{\mathrm{rec}}
=
\|y-\hat{y}\|_1
+
\lambda\,\mathcal{L}_{\mathrm{aux}},
\]
where $\mathcal{L}_{\mathrm{aux}}$ may include perceptual, geometric, or rendering-based terms. Amortized inference enables sub-second generation; generalization is bounded by training data coverage.

\section{Data Foundations and Benchmarks}
\label{sec:data_foundations}
%% Section 3 - Data Foundations and Benchmarks
%% ============================================================
%% Badge helpers (all declared via \providecommand to avoid
%% conflicts with identically-named commands in other sections).
%% ============================================================
\providecommand{\dsbadge}[2]{%
  \begingroup\setlength{\fboxsep}{1pt}%
  \colorbox{#1!12}{\scriptsize\textsf{\textcolor{#1!65!black}{#2}}}\endgroup}

%% Representation
\providecommand{\reprM}{\dsbadge{blue}{M}}
\providecommand{\reprPC}{\dsbadge{orange}{PC}}
\providecommand{\reprV}{\dsbadge{gray}{Vx}}
\providecommand{\reprMV}{\dsbadge{teal}{MV}}
\providecommand{\reprGS}{\dsbadge{violet}{GS}}

%% Annotation
\providecommand{\annTx}{\dsbadge{magenta}{Tx}}
\providecommand{\annPBR}{\dsbadge{green}{PBR}}
\providecommand{\annUV}{\dsbadge{orange}{UV}}
\providecommand{\annSm}{\dsbadge{cyan}{Sm}}
\providecommand{\annPt}{\dsbadge{violet}{Pt}}
\providecommand{\annLy}{\dsbadge{blue}{Ly}}
\providecommand{\annNav}{\dsbadge{teal}{Nav}}
\providecommand{\annPhys}{\dsbadge{red}{Ph}}
\providecommand{\annInst}{\dsbadge{cyan}{Inst}}
\providecommand{\annMat}{\dsbadge{olive}{Mat}}
\providecommand{\annLg}{\dsbadge{purple}{Lg}}

%% Production-ready properties
\providecommand{\grTopo}{\dsbadge{blue}{Topo}}
\providecommand{\grPBRr}{\dsbadge{green}{PBR}}
\providecommand{\grUVr}{\dsbadge{orange}{UV}}
\providecommand{\grNav}{\dsbadge{teal}{Nav}}
\providecommand{\grPhys}{\dsbadge{red}{Phys}}
\providecommand{\grProc}{\dsbadge{violet}{Proc}}
\providecommand{\grPartial}{\dsbadge{gray}{Par.}}

%% Source type
\providecommand{\srcReal}{\dsbadge{red}{Real}}
\providecommand{\srcSynth}{\dsbadge{blue}{CAD}}
\providecommand{\srcWeb}{\dsbadge{cyan}{Web}}
\providecommand{\srcProc}{\dsbadge{violet}{Proc}}

%% Access (compatible with characters_avatars.tex)
\providecommand{\accOpen}{\dsbadge{teal}{O}}
\providecommand{\accReq}{\dsbadge{orange}{R}}
\providecommand{\accCom}{\dsbadge{gray}{C}}

%% Character dataset badge helpers (moved here for unified table)
\providecommand{\databadge}[2]{\begingroup\setlength{\fboxsep}{1pt}\colorbox{#1}{\scriptsize\textsf{#2}}\endgroup}
\providecommand{\dtypeS}{\databadge{red!12}{S}}
\providecommand{\dtypeD}{\databadge{orange!18}{4D}}
\providecommand{\dtypeM}{\databadge{cyan!18}{M}}
\providecommand{\annoMe}{\databadge{blue!10}{Me}}
\providecommand{\annoTx}{\databadge{magenta!12}{Tx}}
\providecommand{\annoS}{\databadge{green!12}{S}}
\providecommand{\annoX}{\databadge{teal!14}{X}}
\providecommand{\annoK}{\databadge{violet!14}{K}}
\providecommand{\annoL}{\databadge{yellow!22}{L}}
\providecommand{\annoA}{\databadge{orange!24}{A}}
\providecommand{\annoG}{\databadge{pink!18}{G}}
\providecommand{\annoC}{\databadge{gray!25}{C}}
\providecommand{\annoRg}{\databadge{gray!18}{Rg}}

%% ============================================================
%% Dataset card layout: colored left-border minipage cards
%% ============================================================
%% Card command: \dscard{border-color}{name}{year}{access-badge}{stats-line}{property-line}
\providecommand{\dscard}[6]{%
  \begin{minipage}[t]{0.31\textwidth}%
    \vspace{0pt}%
    \noindent\textcolor{#1}{\vrule width 2.5pt\hspace{4pt}}%
    \begin{minipage}[t]{0.90\textwidth}%
      \vspace{1.5pt}%
      {\scriptsize\textsf{#2}~\textit{(#3)}\hfill#4}\par%
      \vspace{1pt}%
      {\scriptsize #5}\par%
      {\scriptsize #6}\par%
      \vspace{2pt}%
    \end{minipage}%
    \par\vspace{2pt}%
  \end{minipage}%
}
%% Wide card variant for character datasets (more fields)
\providecommand{\dscardwide}[6]{%
  \begin{minipage}[t]{0.48\textwidth}%
    \vspace{0pt}%
    \noindent\textcolor{#1}{\vrule width 2.5pt\hspace{4pt}}%
    \begin{minipage}[t]{0.92\textwidth}%
      \vspace{1.5pt}%
      {\scriptsize\textsf{#2}~\textit{(#3)}\hfill#4}\par%
      \vspace{1pt}%
      {\scriptsize #5}\par%
      {\scriptsize #6}\par%
      \vspace{2pt}%
    \end{minipage}%
    \par\vspace{2pt}%
  \end{minipage}%
}
\providecommand{\dsnone}{\textcolor{gray}{---}}
\providecommand{\dsatlasheader}[3]{%
  \begingroup\setlength{\fboxsep}{4pt}%
  \colorbox{#1!14}{%
    \parbox{\dimexpr\linewidth-2\fboxsep\relax}{%
      \scriptsize\textsf{\textbf{#2}}\hfill#3}}%
  \endgroup}
\providecommand{\dsatlasentry}[6]{%
  \begingroup\setlength{\fboxsep}{3pt}\setlength{\fboxrule}{0.45pt}%
  \noindent\fcolorbox{#1!65!black}{white}{%
    \parbox{\dimexpr\linewidth-2\fboxsep-2\fboxrule\relax}{%
      {\scriptsize\textsf{\textbf{#2}}~\textit{(#3)}\hfill#4}\par
      {\scriptsize #5}\par
      {\scriptsize #6}}}%
  \endgroup\par\vspace{3pt}}
\providecommand{\dsboardpanel}[4]{%
  \begingroup\setlength{\fboxsep}{4pt}\setlength{\fboxrule}{0.35pt}%
  \fcolorbox{#1!35!black}{#1!6}{%
    \begin{minipage}[t]{\dimexpr0.485\textwidth-2\fboxsep-2\fboxrule\relax}%
      \vspace{0pt}%
      \begingroup\setlength{\fboxsep}{3pt}%
      \colorbox{#1!16}{%
        \parbox{\dimexpr\linewidth-2\fboxsep\relax}{\scriptsize\textsf{\textbf{#2}}}}%
      \endgroup\par\vspace{2pt}%
      #4\par\vspace*{#3}%
    \end{minipage}}%
  \endgroup}
\providecommand{\dsboardrowcontent}[6]{%
  \textcolor{#1!70!black}{\rule{4pt}{4pt}}\hspace{4pt}%
  \begin{minipage}[t]{\dimexpr\linewidth-11pt\relax}%
    {{\fontsize{6.25}{6.55}\selectfont\textsf{\textbf{#2}}~\textit{(#3)}\hfill#4}\par}%
    {{\fontsize{5.9}{6.15}\selectfont #5}\par}%
    {{\fontsize{5.9}{6.15}\selectfont #6}\par}%
  \end{minipage}}
\providecommand{\dsboardrow}[6]{%
  \begingroup\setlength{\fboxsep}{1pt}%
  \noindent\colorbox{white}{%
    \parbox{\dimexpr\linewidth-2\fboxsep\relax}{%
      \dsboardrowcontent{#1}{#2}{#3}{#4}{#5}{#6}}}%
  \endgroup\par\vspace{0.8pt}}
\providecommand{\dsboardrowlast}[6]{%
  \begingroup\setlength{\fboxsep}{1pt}%
  \noindent\colorbox{white}{%
    \parbox{\dimexpr\linewidth-2\fboxsep\relax}{%
      \dsboardrowcontent{#1}{#2}{#3}{#4}{#5}{#6}}}%
  \endgroup\par}

The capability of any generative model is bounded by its training data.  In production-ready 3D asset generation, this constraint is especially
severe: unlike image synthesis, which draws on internet-scale repositories of
billions of samples, 3D corpora remain orders of magnitude smaller and almost
universally lack the structural annotations required by production pipelines.
A dataset that captures surface geometry accurately may still omit the clean manifold
topology, atlas-based UV parameterization, decoupled physically-based rendering (PBR)
material layers, or skeletal correspondences that distinguish a production-ready asset from
a visual approximation.

This section surveys the principal datasets and benchmarks used in production-ready 3D AIGC
research, organized along the three horizontal asset tiers of the taxonomy presented
in Figure~\ref{fig:taxonomy}: general objects and props
(Section~\ref{sec:data_objects}), characters and avatars
(Section~\ref{sec:data_characters}), and scenes and environments
(Section~\ref{sec:data_scenes}).  Table~\ref{tab:unified_datasets} provides a
consolidated overview of all three tiers.  The section concludes with a cross-cutting
analysis of structural gaps that continue to limit progress toward fully automated,
production-grade generation (Section~\ref{sec:data_gaps}).

%% =============================================================
%% EXPERIMENTAL: Unified compact dataset table (to evaluate space savings)
%% =============================================================
\begin{table*}[t]
\setlength{\abovecaptionskip}{2pt}
\setlength{\belowcaptionskip}{2pt}
\caption{Unified dataset landscape for production-ready 3D generation.
  \textbf{Left:} general objects and scenes; \textbf{Right:} characters and avatars.
  \emph{Ann.}\,=\,annotation;  \emph{PR}\,=\,production-ready affordance.}
\label{tab:unified_datasets}
\centering
\begingroup
\setlength{\tabcolsep}{1.8pt}
\renewcommand{\arraystretch}{1.0}
\fontsize{5.5}{6.6}\selectfont
\noindent\rule{\textwidth}{0.5pt}
%%
%% ========== LEFT PANEL: Objects + Scenes ==========
%%
\begin{minipage}[t]{0.495\textwidth}
\centering
\resizebox{\linewidth}{!}{%
\begin{tabular}{@{} l c r l l c @{}}
\rule{0pt}{1.8ex}\textbf{Dataset} & \textbf{Yr} & \textbf{Scale} & \textbf{Fmt.} & \textbf{Ann.\,/\,PR} & \textbf{Ac.} \\
\midrule
\rowcolor{blue!6}
\multicolumn{6}{@{}l}{\textsf{\textbf{Objects} --- CAD / Synthetic}} \\
ShapeNet~\cite{chang2015shapenet}      & '15 & 51K; 55 cat.           & \reprM         & \annSm\,\annTx\,\annUV\,/\,\grPartial     & \accOpen \\
ModelNet~\cite{wu2015modelnet}         & '15 & 128K; 662 cat.         & \reprM\,\reprV & ---\,/\,\grTopo                             & \accOpen \\
ABC~\cite{koch2019abc}                 & '19 & 1M CAD                 & \reprM         & ---\,/\,\grTopo                             & \accOpen \\
PartNet~\cite{mo2019partnet}           & '19 & 27K; 573K parts        & \reprM         & \annSm\,\annPt\,/\,\grPartial               & \accOpen \\
Text2Shape~\cite{chen2018text2shape}   & '18 & 75K text-shape         & \reprV         & \annSm\,\annLg\,/\,---                      & \accOpen \\
3D-FUTURE~\cite{fu2021future}         & '21 & 17K furn.              & \reprM         & \annTx\,\annSm\,/\,\grTopo\,\grUVr          & \accReq \\
\midrule
\rowcolor{cyan!6}
\multicolumn{6}{@{}l}{\textsf{\textbf{Objects} --- Web-Derived}} \\
Thingi10K~\cite{zhou2016thingi10k}    & '16 & 10K                    & \reprM         & ---\,/\,---                                 & \accOpen \\
PhotoShape~\cite{park2018photoshape}   & '18 & 16.2K chairs           & \reprM         & \annTx\,\annPBR\,\annUV\,/\,\grPartial      & \accOpen \\
Objaverse~\cite{deitke2023objaverse}  & '23 & 800K                   & \reprM         & \annTx\,\annUV\,\annPBR\,/\,\grPartial      & \accOpen \\
Cap3D~\cite{luo2023cap3d}             & '23 & 660K text-3D           & \reprM         & \annTx\,\annLg\,/\,\grPartial               & \accOpen \\
Objaverse-XL~\cite{deitke2023objaversexl} & '23 & 10.2M              & \reprM\,\reprMV & \annTx\,/\,\grPartial                      & \accOpen \\
\midrule
\rowcolor{red!6}
\multicolumn{6}{@{}l}{\textsf{\textbf{Objects} --- Real Capture}} \\
CO3D~\cite{reizenstein2021co3d}        & '21 & 1.5M fr.; 19K obj.    & \reprMV\,\reprPC & ---\,/\,---                               & \accOpen \\
GSO~\cite{downs2022google}             & '22 & 1K; 17 cat.            & \reprM         & \annTx\,\annPBR\,\annUV\,/\,\grTopo\,\grPBRr\,\grUVr & \accOpen \\
ABO~\cite{collins2022abo}              & '22 & 8K home goods          & \reprM         & \annTx\,\annPBR\,/\,\grTopo\,\grPBRr        & \accOpen \\
OmniObj3D~\cite{wu2023omniobject3d}    & '23 & 6K; 190 cat.           & \reprM\,\reprPC\,\reprMV & \annTx\,/\,\grPartial          & \accOpen \\
MVImgNet~\cite{yu2023mvimgnet}         & '23 & 6.5M fr.; 238 cat.    & \reprMV\,\reprPC & ---\,/\,---                               & \accOpen \\
uCO3D~\cite{liu2025uco3d}             & '25 & 170K seq.; 1K cat.     & \reprMV\,\reprGS & \annLg\,/\,---                            & \accOpen \\
\midrule
\rowcolor{red!6}
\multicolumn{6}{@{}l}{\textsf{\textbf{Scenes} --- Real Indoor Scans}} \\
ScanNet~\cite{dai2017scannet}          & '17 & 1.5K scans; 2.5M fr.  & RGB-D          & \annSm\,\annInst\,/\,---                     & \accReq \\
ScanNet++~\cite{yeshwanth2023scannetpp}& '23 & 460 scenes             & laser+DSLR     & \annSm\,\annInst\,\annMat\,/\,\grTopo\,\grPBRr & \accReq \\
Matterport3D~\cite{chang2017matterport3d} & '17 & 90 bldgs.          & panorama       & \annSm\,\annInst\,/\,\grTopo                 & \accReq \\
HM3D~\cite{ramakrishnan2021hm3d}       & '21 & 1K buildings           & mesh           & \annSm\,\annNav\,/\,\grNav                   & \accReq \\
\midrule
\rowcolor{blue!6}
\multicolumn{6}{@{}l}{\textsf{\textbf{Scenes} --- Synthetic / Artist}} \\
Structured3D~\cite{zheng2020structured3d} & '20 & 3.5K houses         & render         & \annSm\,\annInst\,\annLy\,/\,\grTopo\,\grUVr & \accOpen \\
Hypersim~\cite{roberts2021hypersim}    & '21 & 461 scenes; 77K img.   & render         & \annSm\,\annInst\,\annMat\,/\,\grPBRr\,\grTopo & \accOpen \\
3D-FRONT~\cite{fu20213d}              & '21 & 18.8K rooms            & artist CAD     & \annSm\,\annLy\,/\,\grTopo\,\grUVr           & \accOpen \\
\midrule
\rowcolor{violet!6}
\multicolumn{6}{@{}l}{\textsf{\textbf{Scenes} --- Procedural / Interactive}} \\
ProcTHOR~\cite{deitke2022procthor}     & '22 & 10K+ houses            & procedural     & \annSm\,\annNav\,\annPhys\,/\,\grNav\,\grPhys\,\grProc & \accOpen \\
Infinigen~\cite{raistrick2023infinite} & '23 & Unlim.\ (outdoor)      & procedural     & \annSm\,\annMat\,/\,\grPBRr\,\grProc         & \accOpen \\
Inf.\ In.~\cite{raistrick2024infinigenindoors} & '24 & Unlim.\ (indoor) & procedural  & \annSm\,\annMat\,\annPhys\,/\,\grPBRr\,\grPhys\,\grProc & \accOpen \\
\end{tabular}}%
\end{minipage}\hfill
%%
%% ========== RIGHT PANEL: Characters ==========
%%
\begin{minipage}[t]{0.495\textwidth}
\centering
\resizebox{\linewidth}{!}{%
\begin{tabular}{@{} l c r l l c @{}}
\rule{0pt}{1.8ex}\textbf{Dataset} & \textbf{Yr} & \textbf{Scale} & \textbf{Type} & \textbf{Ann.\,/\,PR} & \textbf{Ac.} \\
\midrule
\rowcolor{red!6}
\multicolumn{6}{@{}l}{\textsf{\textbf{Characters} --- Static 3D Scans}} \\
FAUST~\cite{bogo2014faust}             & '14 & 300 scans; 10 subj.   & mesh           & \annoMe\,/\,---                              & \accReq \\
RenderPeople~\cite{renderpeople}       & '18 & 4.5K+ subj.           & mesh           & \annoMe\,\annoTx\,\annoRg\,/\,---             & \accCom \\
THuman~\cite{zheng2019deephuman}       & '19 & 7K                    & RGB-D+mesh     & \annoMe\,\annoTx\,\annoS\,/\,---              & \accOpen \\
THuman2.0~\cite{thuman20}              & '21 & 500 subj.             & mesh           & \annoMe\,\annoTx\,\annoX\,/\,---              & \accOpen \\
CTD Static~\cite{chen2021tightcap}     & '21 & 228 garments          & mesh           & \annoMe\,\annoG\,/\,---                        & \accOpen \\
AGORA~\cite{patel2021agora}            & '21 & 14.5K fr.             & synth+real     & \annoX\,\annoTx\,/\,---                        & \accOpen \\
\midrule
\rowcolor{orange!6}
\multicolumn{6}{@{}l}{\textsf{\textbf{Characters} --- Dynamic 4D Capture}} \\
CAPE~\cite{ma2020learning}             & '20 & 15 subj.; 600+ seq.   & 4D mesh        & \annoMe\,\annoS\,/\,---                        & \accReq \\
HuMMan~\cite{cai2022humman}           & '22 & 1K subj.; 60M fr.     & RGB-D+mesh     & \annoMe\,\annoTx\,\annoS\,\annoK\,/\,---       & \accOpen \\
DNA-Rend.~\cite{cheng2023dna}         & '23 & 500 subj.; 5K seq.    & 60-view vid.   & \annoX\,\annoK\,\annoC\,/\,---                 & \accReq \\
X-Humans~\cite{shen2023xavatar}       & '23 & 20 subj.; 35.5K fr.   & textured 4D    & \annoMe\,\annoTx\,\annoX\,/\,---               & \accOpen \\
ActorsHQ~\cite{isik2023humanrf}        & '23 & 16 seq.; 39.8K fr.    & 160-cam vid.   & \annoMe\,\annoC\,/\,---                         & \accReq \\
4D-DRESS~\cite{wang20244ddress}        & '24 & 32 subj.; 520+ seq.   & 4D scans       & \annoMe\,\annoTx\,\annoX\,\annoG\,/\,---        & \accReq \\
\midrule
\rowcolor{magenta!6}
\multicolumn{6}{@{}l}{\textsf{\textbf{Characters} --- Face \& Head}} \\
CoMA~\cite{ranjan2018coma}             & '18 & 12 subj.; 20.5K fr.   & 3D face        & \annoMe\,/\,---                                & \accOpen \\
FFHQ~\cite{karras2019stylegan}        & '19 & 70K images             & 2D face        & ---\,/\,---                                    & \accOpen \\
VoxCeleb~\cite{nagrani2017voxceleb}    & '17 & 1.3K subj.; 153K seq. & audio-visual   & ---\,/\,---                                    & \accOpen \\
CelebV-HQ~\cite{he2022celebvhq}       & '22 & 15.7K subj.; 35.7K seq. & video        & ---\,/\,---                                    & \accOpen \\
VFHQ~\cite{xie2022vfhq}               & '22 & 16K seq.; 1.5M fr.    & HQ video       & ---\,/\,---                                    & \accOpen \\
NeRSemble~\cite{kirschstein2023nersemble} & '23 & 220 subj.          & 16-cam vid.    & \annoMe\,\annoC\,/\,---                         & \accReq \\
\midrule
\rowcolor{cyan!6}
\multicolumn{6}{@{}l}{\textsf{\textbf{Characters} --- Motion Corpora}} \\
Human3.6M~\cite{ionescu2014human36m}   & '14 & 11 subj.; 3.6M fr.    & RGB-D+mocap    & \annoK\,\annoC\,/\,---                          & \accReq \\
AMASS~\cite{mahmood2019amass}          & '19 & 344 subj.; 11.3K seq. & SMPL motion    & \annoS\,/\,---                                  & \accReq \\
HumanML3D~\cite{guo2022humanml3d}     & '22 & 14.6K seq.             & motion+text    & \annoS\,\annoL\,/\,---                          & \accOpen \\
AIST++~\cite{li2021aist}               & '21 & 30 subj.; 1.4K seq.   & dance+music    & \annoS\,\annoK\,\annoA\,\annoC\,/\,---          & \accOpen \\
Motion-X~\cite{lin2023motion}          & '23 & 81.1K seq.; 15.6M fr. & SMPL-X+text    & \annoX\,\annoL\,/\,---                          & \accOpen \\
Motion-X++~\cite{zhang2025motionplus}  & '25 & 120.5K seq.; 19.5M fr.& SMPL-X+RGB     & \annoX\,\annoL\,\annoA\,/\,---                  & \accOpen \\
\end{tabular}}%
\end{minipage}
\par\vspace{-1pt}
\noindent\rule{\textwidth}{0.5pt}
%%
%% ========== LEGEND ==========
%%
\par\vspace{10pt}
\begin{minipage}{\textwidth}
\fontsize{4.9}{5.6}\selectfont
\textbf{Obj format:} \reprM mesh; \reprPC pt.cld; \reprV vox; \reprMV MV; \reprGS 3DGS. \textbf{Obj/Scene ann:} \annTx tex; \annPBR PBR; \annUV UV; \annSm sem; \annPt part; \annLg lang; \annInst inst; \annMat mat; \annLy layout; \annNav nav; \annPhys phys.\\
\textbf{Char ann:} \annoMe mesh; \annoTx tex; \annoS SMPL; \annoX SMPL-X; \annoK kpt; \annoL lang; \annoA audio; \annoG garment; \annoC cam; \annoRg rigged. \textbf{PR:} \grTopo topo; \grPBRr PBR; \grUVr UV; \grPartial partial; \grNav nav; \grPhys phys; \grProc proc.\\
\textbf{Access:} \accOpen open; \accReq reg; \accCom commercial.
\end{minipage}
\endgroup
\vspace{-10pt}
\end{table*}
%% ============= END: Unified compact table =============

%% =========================================================
\subsection{General Objects and Props}
\label{sec:data_objects}
%% =========================================================

Table~\ref{tab:unified_datasets} (left panel) summarizes representative general-object datasets
evaluated from a game-production perspective.

General-object datasets have evolved from compact CAD archives (ShapeNet~\cite{chang2015shapenet}, ModelNet~\cite{wu2015modelnet}, ABC~\cite{koch2019abc}), through internet-scale aggregation (Objaverse~\cite{deitke2023objaverse}, Objaverse-XL~\cite{deitke2023objaversexl}), toward high-fidelity real captures with full PBR material stacks (GSO~\cite{downs2022google}, ABO~\cite{collins2022abo}).
The central tension is a scale--quality trade-off: web-derived collections offer two orders of magnitude more data than curated repositories but contain heterogeneous topology, inconsistent UV maps, and baked lighting artifacts; curated PBR-complete datasets remain too small to train large generative models.
No single collection yet satisfies all four production criteria simultaneously---topological regularity, complete PBR coverage, semantic part decomposition, and language alignment---as Table~\ref{tab:unified_datasets} makes explicit.

%% =========================================================
\subsection{Characters and Avatars}
\label{sec:data_characters}
%% =========================================================

Character and avatar generation imposes the most demanding data requirements among the
three asset tiers, depending on four tightly coupled supervision forms: static body scans for shape and UV structure, dynamic 4D capture for non-rigid deformation, head-specific corpora for expression and speech, and motion datasets for temporal control (Table~\ref{tab:unified_datasets}, right).
Two implications stand out.  First, SMPL and SMPL-X serve not merely as annotations but as operational interfaces: datasets aligned to these parametric models already expose skeletal hierarchies and motion spaces that map naturally to engine rigs.
Second, no single public corpus jointly provides clean body topology, garment deformation, detailed facial dynamics, and language-aligned motion at internet scale, explaining why the character chapter is organized around body synthesis, head synthesis, and rigging.

%% =========================================================
\subsection{Scenes and Environments}
\label{sec:data_scenes}
%% =========================================================

The lower rows of Table~\ref{tab:unified_datasets} (left panel) summarize representative
scene datasets from a game-production perspective.

Scene data presents qualitatively different challenges: a building-scale capture encodes tens of thousands of spatially coherent surface elements requiring appearance, navigability, semantic, collision, and layout annotations.
Real indoor scans (ScanNet~\cite{dai2017scannet}, ScanNet++~\cite{yeshwanth2023scannetpp}, Matterport3D~\cite{chang2017matterport3d}, HM3D~\cite{ramakrishnan2021hm3d}) provide evaluation benchmarks with increasing quality, though topological regularity varies.
Synthetic corpora (Structured3D~\cite{zheng2020structured3d}, Hypersim~\cite{roberts2021hypersim}, 3D-FRONT~\cite{fu20213d}) offer controlled conditions with complete material decomposition and compositional layout annotations.
Procedural generators (ProcTHOR~\cite{deitke2022procthor}, Infinigen~\cite{raistrick2023infinite, raistrick2024infinigenindoors}) provide effectively unlimited scale with full collision, navigation, and material annotations; Infinigen Indoors offers direct Unreal Engine 5 export.

%% =========================================================
\subsection{Structural Gaps and Implications for Production-Ready Generation}
\label{sec:data_gaps}
%% =========================================================

The preceding survey reveals four structural gaps that collectively constrain the
quality ceiling of production-ready generation models.

\paragraph{Absence of topology-quality supervision.}
No large-scale 3D dataset provides explicit annotations for production topology quality---face-aligned edge loops, quad-dominant structure, or density gradients.  ShapeNet, Objaverse, and derivatives contain substantial topologically irregular meshes unsuitable for rigging, UV parameterization, or real-time rendering.  Without ground-truth topology labels, neither supervised retopology nor evaluation metrics for topological fidelity can be established.

\paragraph{Limited coverage of PBR material annotations.}
Only GSO~\cite{downs2022google}, ABO~\cite{collins2022abo}, ScanNet++~\cite{yeshwanth2023scannetpp}, Hypersim~\cite{roberts2021hypersim}, and Infinigen Indoors~\cite{raistrick2024infinigenindoors} provide complete PBR channels (albedo, normal, roughness, metallic, AO).  Their combined scale falls far short of Objaverse-XL's 10.2M instances~\cite{deitke2023objaversexl}, preventing PBR estimation and geometry generation from being co-trained at internet scale.

\paragraph{Cross-asset consistency and style coherence.}
Datasets for objects, characters, and scenes were developed independently with different scale conventions, coordinate frames, and material standards.  No large-scale corpus jointly annotates assets across tiers with coherent lighting, style descriptors, or compatible material parameters, preventing training of models that produce style-consistent scene populations---a fundamental requirement for game-level art direction.

\paragraph{Absence of standardized game-readiness benchmarks.}
Common metrics borrowed from image generation (FID, CLIP similarity) assess appearance fidelity but are silent on topology validity, UV-seam density, PBR accuracy, and engine import success rate.  The absence of an agreed-upon production-readiness evaluation suite makes it difficult to quantify systematically the production gap between methods or track progress on deployment-critical properties.  This gap is identified as a primary open challenge in Section~\ref{sec:conclusion}.

\section{General Objects and Props}
\label{sec:general_objects}
General objects span a wide range of shape complexity, posing substantial challenges for automated generation. Production pipelines additionally require that generated assets conform to technical standards, including clean topology~\cite{liu2025quadgpt}, semantically coherent UV unwrapping~\cite{li2025auto}, and physically based material properties~\cite{zhao2025hunyuan3d, he2025materialmvp}. Having reviewed data foundations in Section~\ref{sec:data_objects}, this section focuses on how method families address these requirements along the production pipeline.

We organize this chapter according to the industrial asset production workflow:
\begin{itemize}
    \item \textbf{Geometry Generation}: Synthesizing coarse 3D shapes from user inputs. Outputs at this stage typically lack clean topology, exhibiting inconsistent connectivity, dense triangulations, or surface holes.

    \item \textbf{Topology Generation}: Restructuring raw geometry into artist-friendly mesh connectivity that supports real-time rendering and further downstream processing.

    \item \textbf{Appearance Generation}: Establishing a bijective UV mapping between the 3D surface and a 2D parameter domain, and synthesizing physically based surface materials, including albedo, normal, roughness, metallicity, and displacement maps---to ensure PBR consistency with real-time game engines.
\end{itemize}

\subsection{Geometry Generation}

Geometry generation synthesizes coarse shapes from user inputs, providing editable initial geometries for subsequent topology optimization, UV unwrapping, and texture mapping~\cite{lei2025hunyuan3d}. Two paradigms have emerged: direct training on large-scale 3D data~\cite{wu2016learning, gao2019sdm, cai2020learning}, and distillation from 2D generative models without 3D supervision~\cite{poole2023dreamfusion, lin2023magic3d, chen2023fantasia3d}.

\begin{table*}[t]
\caption{Typical representative geometry generation methods for general objects and props.}
\label{tab:geometry_generation_methods}
\centering
\begin{threeparttable}
\fontsize{6}{7.2}\selectfont
\setlength{\tabcolsep}{2.5pt}
\renewcommand{\arraystretch}{1.08}
\begin{tabularx}{\textwidth}{@{} l c l l X X l @{}}
\toprule
\rowcolor{colHead}
\textcolor{white}{\textbf{Method}} &
\textcolor{white}{\textbf{Year}} &
\textcolor{white}{\textbf{Input}} &
\textcolor{white}{\textbf{Representation}} &
\textcolor{white}{\textbf{Approach}} &
\textcolor{white}{\textbf{Highlight}} &
\textcolor{white}{\textbf{Group}} \\
\midrule
%% 2D Prior Driven
\rowcolor{grpSdsRow}
DreamFusion~\cite{poole2023dreamfusion}       & 2022 & Text         & NeRF                & SDS                 & Open-domain        & \cellcolor{grpSdsHead}\textcolor{white}{\textsf{SDS}} \\
Magic3D~\cite{lin2023magic3d}                 & 2023 & Text         & NeRF $\to$ mesh     & Coarse-to-fine SDS  & Finer detail       & \cellcolor{grpSdsRow}\textsf{SDS} \\
\rowcolor{grpSdsRow}
Fantasia3D~\cite{chen2023fantasia3d}          & 2023 & Text         & DMTet               & Disentangled SDS    & Textured mesh      & \textsf{SDS} \\
ProlificDreamer~\cite{wang2023prolificdreamer}& 2023 & Text         & NeRF                & Variational SDS     & Less over-smooth   & \cellcolor{grpSdsRow}\textsf{SDS} \\
\rowcolor{grpSdsRow}
RichDreamer~\cite{qiu2024richdreamer}         & 2024 & Text         & Implicit            & Normal-depth prior  & Stable geometry    & \textsf{SDS} \\
\rowcolor{grpMvRow}
Zero-1-to-3~\cite{liu2023zero}                & 2023 & Single image & NeRF                & View-cond.\ diff.   & Novel-view         & \cellcolor{grpMvHead}\textcolor{white}{\textsf{MV}} \\
MVDream~\cite{shi2023MVDream}                 & 2024 & Text / image & Agnostic            & MV diffusion        & MV consistency     & \cellcolor{grpMvRow}\textsf{MV} \\
\rowcolor{grpMvRow}
Wonder3D~\cite{long2024wonder3d}              & 2024 & Single image & Mesh                & Color+normal diff.  & Normal-guided      & \textsf{MV} \\
SV3D~\cite{voleti2024sv3d}                    & 2024 & Single image & Mesh / 3DGS         & Video diff.         & Dense MV           & \cellcolor{grpMvRow}\textsf{MV} \\
\midrule
%% 3D Data Driven
\rowcolor{grpGanRow}
3D-GAN~\cite{wu2016learning}                  & 2016 & Noise        & Voxel               & Adversarial         & Pioneering         & \cellcolor{grpGanHead}\textcolor{white}{\textsf{GAN}} \\
Tree-GAN~\cite{shu20193d}                     & 2019 & Noise        & Point cloud         & Tree-struct.\ GAN   & Structural pts     & \cellcolor{grpGanRow}\textsf{GAN} \\
\rowcolor{grpGanRow}
SP-GAN~\cite{li2021sp}                        & 2021 & Noise        & Point cloud         & Spherical prior     & Global prior       & \textsf{GAN} \\
SDF-StyleGAN~\cite{zheng2022sdfstylegan}      & 2022 & Noise        & SDF                 & StyleGAN on SDF     & High-res SDF       & \cellcolor{grpGanRow}\textsf{GAN} \\
\rowcolor{grpVaeRow}
AtlasNet~\cite{groueix2018papier}             & 2018 & Image/latent & Patches             & Patch deform.\ VAE  & Surface recon.     & \cellcolor{grpVaeHead}\textcolor{white}{\textsf{VAE}} \\
TM-Net~\cite{gao2021tm}                       & 2021 & Image/latent & Mesh + texture      & Geom.-tex.\ VAE     & Texture-ready      & \cellcolor{grpVaeRow}\textsf{VAE} \\
\rowcolor{grpVaeRow}
Michelangelo~\cite{zhao2023michelangelo}      & 2023 & Text / image & Latent shape        & Aligned VAE         & Multimodal         & \textsf{VAE} \\
CLAY~\cite{zhang2024clay}                     & 2024 & Text/img/ctrl& Latent 3D           & VAE + DiT           & Controllable       & \cellcolor{grpVaeRow}\textsf{VAE} \\
\rowcolor{grpDdmRow}
PC-DPM~\cite{luo2021diffusion}                & 2021 & Noise        & Point cloud         & DDPM                & Point denoising    & \cellcolor{grpDdmHead}\textcolor{white}{\textsf{Diff}} \\
MeshDiffusion~\cite{liu2023meshdiffusion}     & 2023 & Noise        & Mesh                & Score diffusion     & Direct mesh        & \cellcolor{grpDdmRow}\textsf{Diff} \\
\rowcolor{grpDdmRow}
TetraDiffusion~\cite{kalischek2024tetradiffusion}& 2024 & Text/image & Tetrahedral         & Tetra.\ diffusion   & High-res topo.     & \textsf{Diff} \\
\rowcolor{grpFfRow}
Pixel2Mesh~\cite{wang2018pixel2mesh}          & 2018 & Single image & Deform.\ mesh       & GCN deform.         & Direct mesh        & \cellcolor{grpFfHead}\textcolor{white}{\textsf{FF}} \\
LRM~\cite{hong2024lrm}                        & 2023 & Single image & Tri-plane           & Transformer LRM     & Fast recon.        & \cellcolor{grpFfRow}\textsf{FF} \\
\rowcolor{grpFfRow}
TripoSR~\cite{TripoSR2024}                    & 2024 & Single image & Implicit / mesh     & Distilled FF        & Sub-second         & \textsf{FF} \\
InstantMesh~\cite{xu2024instantmesh}          & 2024 & MV images$^\dagger$  & FlexiCubes          & MV + mesh pred.     & Mesh + UV          & \cellcolor{grpFfRow}\textsf{FF} \\
\rowcolor{grpFfRow}
SF3D~\cite{sf3d2024}                          & 2024 & Single image & Mesh + UV + PBR     & FF + material       & PBR-ready          & \textsf{FF} \\
Fast3R~\cite{yang2025fast3r}                  & 2025 & Multi-view   & 3D recon.           & Amortized           & Scalable           & \cellcolor{grpFfRow}\textsf{FF} \\
\rowcolor{grpLgmRow}
Shap-E~\cite{jun2023shap}                     & 2023 & Text / image & Implicit / mesh     & Latent diff.        & Fast gen.          & \cellcolor{grpLgmHead}\textcolor{white}{\textsf{LGM}} \\
3DShape2VecSet~\cite{zhang20233dshape2vecset} & 2023 & Latent       & Vector-set          & Vec-set VAE+diff.   & Unordered latents  & \cellcolor{grpLgmRow}\textsf{LGM} \\
\rowcolor{grpLgmRow}
XCube~\cite{ren2024xcube}                     & 2024 & Text/img/ctrl& Sparse voxel        & Hierarchical latent & Large-scale        & \textsf{LGM} \\
TRELLIS~\cite{xiang2025structured}            & 2025 & Text / image & SLAT                & Rect.\ flow         & Mesh / 3DGS        & \cellcolor{grpLgmRow}\textsf{LGM} \\
TRELLIS.2~\cite{xiang2025trellis2}           & 2025 & Image        & O-Voxel             & Rect.\ flow         & 4B params, PBR     & \cellcolor{grpLgmRow}\textsf{LGM} \\
\rowcolor{grpLgmRow}
SparseFlex~\cite{he2025sparseflex}            & 2025 & Latent/cond. & Sparse isosurface   & VAE + flow          & Arb.\ topology     & \textsf{LGM} \\
TripoSG~\cite{li2025triposg}                  & 2025 & Image        & Latent tokens       & VAE + rect.\ flow   & High-fidelity      & \cellcolor{grpLgmRow}\textsf{LGM} \\
\rowcolor{grpLgmRow}
MeshCraft~\cite{he2025meshcraft}              & 2025 & Image/latent & Face tokens         & VAE + flow DiT      & Parallel gen.      & \textsf{LGM} \\
\bottomrule
\end{tabularx}
\smallskip
\begin{tablenotes}[flushleft]\fontsize{5.5}{6.5}\selectfont
\item[] \textit{Group labels\,:}\;
  \textsf{SDS}\,=\,Score Distillation;\;
  \textsf{MV}\,=\,Multi-view Recon.;\;
  \textsf{GAN};\;
  \textsf{VAE}\,=\,VAE/AE;\;
  \textsf{Diff}\,=\,Direct Diffusion;\;
  \textsf{FF}\,=\,Feed-Forward;\;
  \textsf{LGM}\,=\,Latent Gen.
  First row of each group is highlighted with the group color.
\item[$\dagger$] Trained on 3D data; MV images at inference are typically generated by a 2D diffusion prior.
\end{tablenotes}
\end{threeparttable}
\vspace{-10pt}
\end{table*}

\textbf{Methods driven by 2D priors.}  This family transfers knowledge from pretrained 2D models to a 3D representation $\mathbf{z}$ through differentiable rendering $\mathcal{R}$:
\begin{equation*}
\min_{\mathbf{z}}\;\mathcal{L}_{\mathrm{2D}}\!\bigl(\mathcal{R}(\mathbf{z}),\;\mathcal{P}_{\mathrm{2D}}\bigr),
\end{equation*}
where $\mathcal{P}_{\mathrm{2D}}$ denotes a 2D prior. Two instantiations dominate: \emph{per-scene score distillation}~\cite{poole2023dreamfusion,lin2023magic3d,chen2023fantasia3d,wang2023prolificdreamer,qiu2024richdreamer,tang2023dreamgaussian,yi2023gaussiandreamer} and \emph{multi-view generation followed by reconstruction}~\cite{liu2023zero,shi2023MVDream,long2024wonder3d,voleti2024sv3d}. These methods excel at open-domain semantics but do not directly optimize topology, UV, or rigging readiness (Table~\ref{tab:geometry_generation_methods}).

\textbf{Methods driven by 3D data.}
Under 3D data supervision, geometry generation learns a mapping $\hat{x} = G_{\theta}(c,\xi)$ into a target 3D space $\mathcal{X}$, optimized via paradigm-specific losses $\mathcal{L}$. The supervision frontier has expanded from ShapeNet~\cite{chang2015shapenet} to Objaverse-XL~\cite{deitke2023objaverse,deitke2023objaversexl}. We organize these methods into five paradigms:

\paragraph{GANs, VAEs, and Direct Diffusion.}
3D GANs progressed from voxels~\cite{wu2016learning} through point clouds~\cite{achlioptas2018learning, shu20193d, li2021sp} to implicit fields~\cite{zheng2022sdfstylegan}, while VAE/AE methods evolved from patch warping~\cite{groueix2018papier} to joint geometry-texture generation~\cite{gao2021tm}. Direct 3D diffusion advanced from point clouds~\cite{luo2021diffusion, zhou20213d, nichol2022point} to meshes~\cite{liu2023meshdiffusion} and tetrahedral partitions~\cite{kalischek2024tetradiffusion}, but computational cost and limited topology control have driven adoption of latent frameworks.

\paragraph{Feed-Forward Reconstruction Models.}
Feed-forward models amortize 3D prediction from sparse observations for interactive-speed generation. LRM~\cite{hong2024lrm} and TripoSR~\cite{TripoSR2024} achieve sub-second single-image reconstruction, GS-LRM~\cite{zhang2024gs} extends to Gaussians, and multi-view pipelines (InstantMesh~\cite{xu2024instantmesh}, LGM~\cite{tang2024lgm}) further improve quality. SF3D~\cite{sf3d2024} jointly predicts geometry, UV, and PBR materials, representing the closest approach to production-ready feed-forward generation.

\paragraph{Latent Generative Models.}
Latent generative models operate in compact learned spaces using diffusion~\cite{ho2020denoising, ren2024xcube, zhao2023michelangelo, zhang2024clay} or rectified flow~\cite{liu2022flow, xiang2025structured, xiang2025trellis2, he2025sparseflex}. Two dominant organizations have emerged: \textbf{sparse structured spatial latents}---XCube~\cite{ren2024xcube}, TRELLIS~\cite{xiang2025structured} and its successor TRELLIS.2~\cite{xiang2025trellis2}, SparseFlex~\cite{he2025sparseflex}---and \textbf{vector-set latents}---3DShape2VecSet~\cite{zhang20233dshape2vecset}, Michelangelo~\cite{zhao2023michelangelo}, CLAY~\cite{zhang2024clay}, LATTICE~\cite{lai2025lattice}. TRELLIS.2 advances this line with an \emph{O-Voxel} representation that supports non-manifold geometry and transparent surfaces while scaling to 4B parameters. The latest systems TripoSG~\cite{li2025triposg} and MeshCraft~\cite{he2025meshcraft} push toward high-fidelity mesh generation via flow-based DiT architectures.

\textbf{Part-aware geometry organization.}
Production assets require per-part control for downstream topology and deformation. Two directions have emerged: \emph{recovering parts from holistic geometry}---SAMPart3D~\cite{yang2024sampart3d} for multi-granularity segmentation, HoloPart~\cite{yang2025holopart} for amodal completion, X-Part~\cite{yan2025x} for controllable decomposition---and \emph{generating geometry at the part level}---PAGENet~\cite{li2020learning}, PartGen~\cite{chen2025partgen}, PartCrafter~\cite{lin2025partcrafter}, and OmniPart~\cite{yang2025omnipart} for part-wise reconstruction and editing.

\begin{figure*}[t]
    \centering
    \includegraphics[width=\textwidth]{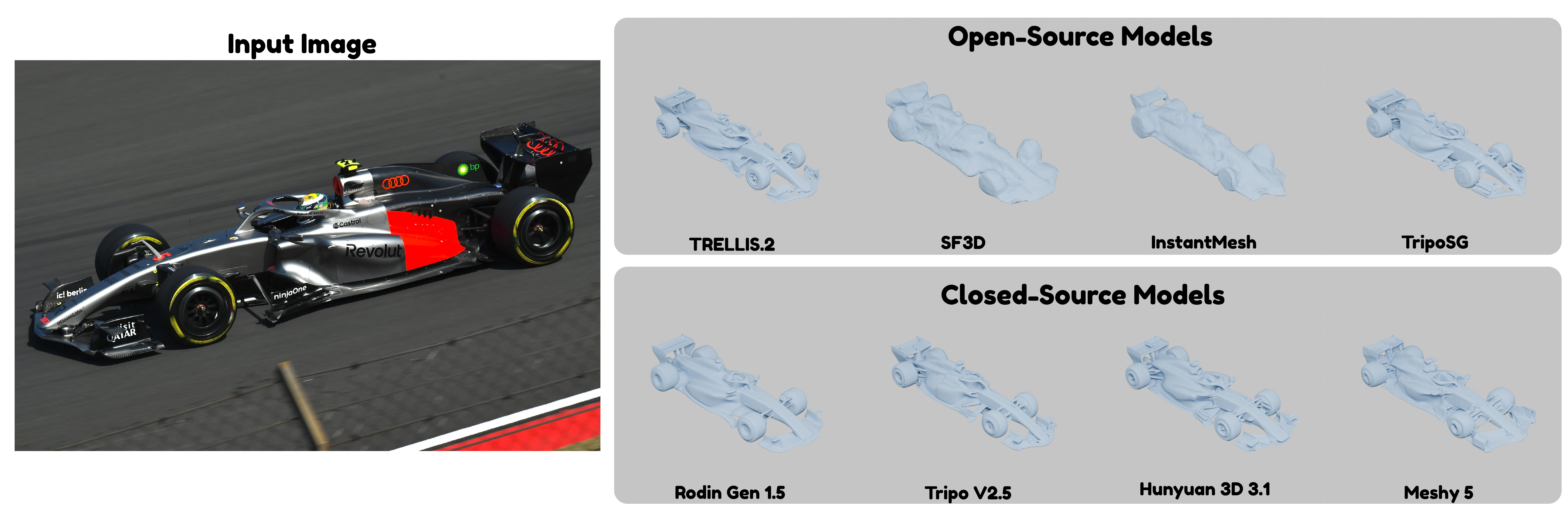}
    \caption{Single-image 3D reconstruction results from open-source and closed-source models on the same input photograph.
    Top row (open-source): TRELLIS.2~\cite{xiang2025trellis2}, SF3D~\cite{sf3d2024}, InstantMesh~\cite{xu2024instantmesh}, TripoSG~\cite{li2025triposg}.
    Bottom row (closed-source): Rodin Gen~1.5~\cite{rodinGen2025}, Tripo~V2.5~\cite{tripoV25_2025}, Hunyuan3D~3.1~\cite{hunyuan3d31_2025}, Meshy~5~\cite{meshy5_2025}.
    Closed-source systems generally achieve superior geometric fidelity and surface detail over their open-source counterparts.}
    \Description{A side-by-side comparison of 3D reconstructions from a single F1 car photograph. The left panel shows the input image. The right panel contains two rows of clay-rendered 3D meshes: the top row shows results from four open-source models (TRELLIS.2, SF3D, InstantMesh, TripoSG) and the bottom row shows results from four closed-source commercial systems (Rodin Gen 1.5, Tripo V2.5, Hunyuan3D 3.1, Meshy 5).}
    \label{fig:open_closed_comparison}
    \vspace{-8pt}
\end{figure*}

\subsection{Topology Generation}
Geometry generation often produces dense triangle soups with irregular connectivity, inconsistent face density, and deformation-unaware edge flow. Topology generation reformulates these into meshes compatible with editing, UV parameterization, rigging, and engine deployment. Existing methods fall into two categories: \emph{indirect} methods that treat topology as post-hoc optimization over existing geometry~\cite{instant2015field, exoside2019quadremesher}, and \emph{direct} methods that generate artist-aligned connectivity end-to-end~\cite{siddiqui2024meshgpt, liu2025quadgpt}.

\textbf{Indirect methods.}
Classical remeshing---quadric simplification~\cite{10.1145/258734.258849}, Instant Meshes~\cite{instant2015field}, QuadriFlow~\cite{huang2018quadriflow}---remains the production baseline, with learning-based extensions~\cite{Potamias_2022_CVPR, dong2025neurcross}. These methods cannot guarantee deformation-aware edge flow.

\textbf{Direct methods.}
Direct methods generate topology as a primary target. Two routes dominate: \emph{autoregressive generation}, where mesh structure is tokenized and predicted sequentially~\cite{nash2020polygen, siddiqui2024meshgpt}, and \emph{diffusion-based generation}, where topology is produced via iterative denoising.~\cite{alliegro2023polydiff, he2025meshcraft} Autoregressive methods currently dominate, as discretized mesh tokens form structured sequences with strong prefix dependencies, whereas irregular connectivity is inherently harder to corrupt and denoise robustly.

\paragraph{Autoregressive-based generation.}
Autoregressive methods cast topology synthesis as next-token prediction. Two tokenization strategies dominate: \emph{learned neural compression}---MeshGPT~\cite{siddiqui2024meshgpt} introduced VQ-VAE codebooks, extended by shape-conditioned extraction~\cite{chen2024meshanything}, coarse-to-fine scaffolding~\cite{weng2024pivotmesh}, and hybrid AR--latent pipelines~\cite{tang2024edgerunner}---and \emph{geometry-aware serialization}: PolyGen~\cite{nash2020polygen} established vertex-then-face prediction, refined by adjacency compression~\cite{chen2025meshanything}, blocked patchification~\cite{weng2025scaling}, and interleaved attention~\cite{wang2025iflame}. Cross-cutting advances include preference-based alignment (DeepMesh~\cite{zhao2025deepmesh}, Mesh-RFT~\cite{liu2025mesh}) and quad-mesh extension (QuadGPT~\cite{liu2025quadgpt}).

\paragraph{Diffusion-based generation.}
Diffusion-based direct topology generation remains less explored than autoregressive methods, as corrupting and denoising irregular polygonal connectivity without destroying mesh validity is considerably harder than prefix-conditioned sequence prediction~\cite{alliegro2023polydiff, he2025meshcraft}. PolyDiff~\cite{alliegro2023polydiff} formulates this as discrete denoising diffusion over quantized triangle soups, jointly modeling vertex geometry and face connectivity. MeshCraft~\cite{he2025meshcraft} instead encodes meshes into continuous face-level tokens via a transformer-based VAE and generates full topology in parallel through a flow-based DiT with face-count control. SpaceMesh~\cite{shen2024spacemesh} departs from discrete tokenization entirely, representing connectivity as continuous halfedge latents: vertex positions are generated via diffusion, and manifold face structure is recovered by a Sinkhorn-normalized connectivity network, achieving speeds orders of magnitude faster than autoregressive methods. Diffusion approaches offer parallel decoding and global structural refinement, though the literature remains sparse relative to autoregressive formulations.

\paragraph{Artistic alignment.}
Artistic alignment shifts the objective from geometric validity to artist-quality structure: meaningful edge flow, deformation-aware loops, and controllable density. Data imitation approaches---MeshAnything~\cite{chen2024meshanything, chen2025meshanything} and Meshtron~\cite{hao2024meshtron}---train on artist-created meshes but do not explicitly optimize non-local topological qualities. Preference-based methods address this directly: DeepMesh~\cite{zhao2025deepmesh} incorporates RL-based feedback, Mesh-RFT~\cite{liu2025mesh} introduces masked DPO for localized defect correction, and QuadGPT~\cite{liu2025quadgpt} extends alignment to the quad domain for structured edge loops.

\subsection{Appearance Generation}
Appearance generation determines whether an asset can be rendered, edited, and reused in production, encompassing PBR surface attributes---albedo, normal, roughness, and metallic channels---compatible with real-time shading models. The stage couples two sub-problems: \emph{UV unwrapping}, which maps the 3D surface to a 2D parameter domain, and \emph{texture generation}, which synthesizes appearance over that domain. Recent systems such as SF3D~\cite{sf3d2024}, 3DTopia-XL~\cite{chen2025primx}, and MaterialMVP~\cite{he2025materialmvp} increasingly treat UV layout and PBR decomposition as joint objectives, indicating a shift toward relightable, physically grounded assets.

\paragraph{UV unwrapping.}
UV unwrapping maps a 3D surface onto a 2D domain for texture authoring. Classical tools xatlas~\cite{xatlas2018} and UVAtlas~\cite{MicrosoftUVAtlas2023} remain production baselines. Neural methods address their limitations through learned seam prediction~\cite{li2025auto}, unsupervised cycle mapping~\cite{zhang2024flatten, zhao2025flexpara}, and production-oriented objectives including semantic chart alignment~\cite{wang2025partuv}, artist-style packing~\cite{chen2025artuv}, and seam preference optimization~\cite{xu2025seamcrafter}. End-to-end systems such as SF3D~\cite{sf3d2024} and Hunyuan3D Studio~\cite{lei2025hunyuan3d} treat UV construction as integral to reconstruction.

\paragraph{Texture generation.}
Texture generation assigns visual and material information to the UV domain, targeting PBR channels (albedo, normal, roughness, metallic) beyond mere RGB~\cite{deng2024flashtex, 10.1145/3658170}. Three directions have emerged. \emph{Multi-view-consistent synthesis} progressed from iterative inpainting~\cite{richardson2023texture, chen2023text2tex} through cross-view aggregation~\cite{cao2023texfusion, cheng2025mvpaint} to feed-forward UV-space diffusion~\cite{yu2024texgen}. \emph{PBR material generation} targets illumination disentanglement: Paint3D~\cite{zeng2024paint3d} trains illumination-free texture diffusion, and recent methods make PBR decomposition a first-class objective~\cite{he2025materialmvp, huang2024materialanything}. \emph{Unified pipelines}---SF3D~\cite{sf3d2024}, Meta 3D AssetGen~\cite{siddiqui2024assetgen}, PBR3DGen~\cite{wei2025pbr3dgenvlmguidedmeshgeneration}---jointly address UV, geometry, and material. Overall, the field is converging toward multi-view-consistent, UV-aware, and physically grounded material generation.

\section{Characters and Avatars}
\label{sec:characters_and_avatars}
In game development, characters and avatars constitute the most complex class of assets, as they necessitate a balance of artistic aesthetics, anatomical correctness, and functional interactivity. Unlike static props, a character prepared for engine deployment must satisfy strict operational requirements: canonical topology for predictable deformation, hierarchical skeletal rigging for animation control, and physically plausible skinning weights to prevent visual artifacts such as the candy-wrapper effect during motion.

Recent advancements in 2024 and 2025 have shifted the focus from reconstruction to generation. While earlier studies concentrated on the recovery of static geometry from images~\cite{saito2019pifu,alldieck2019tex2shape}, current research, driven by the demands of the metaverse and high-fidelity gaming, aims to directly synthesize animatable entities. This capability encompasses the generation of full-body avatars from single-shot inputs that function immediately within standard graphics pipelines~\cite{liao2024tada,qiu2025lhm}, in addition to the integration of highly expressive features, including physics-aware clothing~\cite{zheng2024physavatar,li2025simavatar} and micro-expressions~\cite{zhang2025motionplus,xu2023omniavatar}.

We categorize the landscape of character generation based on the lifecycle of an asset within a game engine. Because the shared dataset survey has already been consolidated in Section~\ref{sec:data_characters}, the discussion here begins with the structural priors that convert those datasets into controllable generative spaces. It then turns to full-body synthesis, wherein methods generate the geometry and appearance required for 4D animatable avatars, and finally to the rigging bridge that translates raw outputs into articulated assets prepared for engine integration.

\subsection{Structural Priors for Animatable Characters}
\label{sec:char_data}

The structural priors underpinning character generation derive from the datasets consolidated in Section~\ref{sec:data_characters} and Table~\ref{tab:unified_datasets}. Parametric body models---SMPL~\cite{loper2015smpl} and SMPL-X~\cite{pavlakos2019expressive}---provide the most important such prior, converting raw scans into low-dimensional shape, pose, and articulation spaces whose skeletal hierarchies map naturally to engine rigs. Temporal correspondence from 4D capture and motion corpora supplies the second essential signal, encoding cloth dynamics and speech-synchronized motion that static scans cannot capture. These two priors explain why the subsequent discussion separates full-body synthesis, head animation, and rigging into distinct problems.

%% Shared badge helper for the method tables below.
\providecommand{\databadge}[2]{\begingroup\setlength{\fboxsep}{1pt}\colorbox{#1}{\scriptsize\textsf{#2}}\endgroup}

\subsection{Full-Body Geometry and Appearance Synthesis}
\label{sec:full_body_synthesis}

Synthesizing 3D geometry and surface appearance for full-body characters is not merely a reconstruction problem. From a game-production perspective, a useful method must balance at least three requirements that are often in conflict: animatability under a stable skeletal prior, support for free-topology elements such as loose garments and hair, and appearance consistency for unseen regions and viewpoints. The progression of the field can be read as a sequence of attempts to relax one bottleneck without losing the advantages of the previous representation.

% Deliverable badge commands
\providecommand{\catAM}{\databadge{teal!22}{AM}}
\providecommand{\catGS}{\databadge{cyan!22}{GS}}
\providecommand{\catEM}{\databadge{orange!22}{EM}}
\providecommand{\catRO}{\databadge{red!16}{RO}}
\providecommand{\catVD}{\databadge{purple!16}{VD}}
\providecommand{\catAN}{\databadge{blue!20}{AN}}
\providecommand{\catGN}{\databadge{orange!22}{GN}}
% Full-body generation comparison table

%% Table 3: Full-Body Generation Methods
\begin{table*}[t]
\centering
\caption{Full-body character generation methods. Methods are color-coded by group; see legend below.}
\label{tab:full_body_generation}
\fontsize{6}{7.2}\selectfont
\begin{threeparttable}
\setlength{\tabcolsep}{2pt}
\renewcommand{\arraystretch}{0.95}
\begin{tabularx}{\linewidth}{@{} >{\raggedright\arraybackslash}p{0.16\linewidth}
  >{\centering\arraybackslash}p{0.03\linewidth}
  >{\raggedright\arraybackslash}p{0.09\linewidth}
  >{\raggedright\arraybackslash}X
  >{\raggedright\arraybackslash}p{0.10\linewidth}
  >{\centering\arraybackslash}p{0.045\linewidth}
  >{\centering\arraybackslash}p{0.055\linewidth} @{}}
\toprule
\rowcolor{colHead}
\textcolor{white}{\textbf{Method}} &
\textcolor{white}{\textbf{Yr}} &
\textcolor{white}{\textbf{Input}} &
\textcolor{white}{\textbf{3D Representation}} &
\textcolor{white}{\textbf{Prior}} &
\textcolor{white}{\textbf{Del.}} &
\textcolor{white}{\textbf{Group}} \\
\midrule
%% Parametric Template
\rowcolor{grpParamRow}
Tex2Shape~\cite{alldieck2019tex2shape}    & '19 & Img.      & SMPL + Displacements    & 3D/Paired    & \catAM\tnote{a} & \cellcolor{grpParamHead}\textcolor{white}{\textsf{Param}} \\
CAPE~\cite{ma2020learning}                & '20 & 3D Scans  & Mesh (4D Clothing)      & 3D Sup.      & \catAM           & \cellcolor{grpParamRow}\textsf{Param} \\
\rowcolor{grpParamRow}
ExPose~\cite{choutas2020monocular}        & '20 & Img.      & SMPL-X Parameters       & 3D/2D Mix    & \catAM           & \textsf{Param} \\
STAR~\cite{osman2020star}                 & '20 & Img./Vid. & Sparse SMPL Variant     & 3D/Pose      & \catAM           & \cellcolor{grpParamRow}\textsf{Param} \\
\rowcolor{grpParamRow}
HybrIK~\cite{li2021hybrik}               & '21 & Img.      & Skeleton + Mesh         & 2D/3D Mix    & \catAM\tnote{d}  & \textsf{Param} \\
%% Implicit/Hybrid
\rowcolor{grpImplRow}
PIFu~\cite{saito2019pifu}                & '19 & Img.      & Pixel-Aligned Implicit  & 2D/3D Sup.   & \catRO\tnote{a}  & \cellcolor{grpImplHead}\textcolor{white}{\textsf{Impl}} \\
ARCH~\cite{huang2020arch}                & '20 & Img.      & Canon.\ Implicit + Rig  & SMPL+2D/3D   & \catAM           & \cellcolor{grpImplRow}\textsf{Impl} \\
\rowcolor{grpImplRow}
PIFuHD~\cite{saito2020pifuhd}            & '20 & Img.      & Multi-level Implicit    & 2D/3D Sup.   & \catRO\tnote{a}  & \textsf{Impl} \\
PaMIR~\cite{zheng2021pamir}              & '21 & Img.      & Implicit + SMPL         & 2D/3D Weak   & \catRO\tnote{a}  & \cellcolor{grpImplRow}\textsf{Impl} \\
\rowcolor{grpImplRow}
SMPLicit~\cite{corona2021smplicit}        & '21 & Params.   & Implicit Clothing       & 3D Sup.      & \catRO\tnote{a}  & \textsf{Impl} \\
ICON~\cite{xiu2022icon}                  & '22 & Img.      & Implicit + SMPL         & 2D+Normals   & \catRO\tnote{a}  & \cellcolor{grpImplRow}\textsf{Impl} \\
\rowcolor{grpImplRow}
gDNA~\cite{chen2022gdna}                 & '22 & Multi-view & Generative Implicit    & 3D/Weak      & \catRO\tnote{a}  & \textsf{Impl} \\
ECON~\cite{xiu2023econ}                  & '23 & Img.      & Implicit + Mesh         & 2D+Normals   & \catRO\tnote{a}  & \cellcolor{grpImplRow}\textsf{Impl} \\
\rowcolor{grpImplRow}
S3F~\cite{corona2023structured3d}         & '23 & Img.      & Struct.\ 3D Features    & Semi+SMPL    & \catAM           & \textsf{Impl} \\
%% Generative (GAN/Diff)
\rowcolor{grpGenRow}
StylePeople~\cite{grigorev2021stylepeople}& '21 & Img.      & Clothed Mesh            & 2D GAN       & \catEM           & \cellcolor{grpGenHead}\textcolor{white}{\textsf{Gen}} \\
AvatarGen~\cite{zhang2022avatargen}       & '22 & Text/Img. & SDF + Tri-plane         & GAN Prior    & \catEM\tnote{a}  & \cellcolor{grpGenRow}\textsf{Gen} \\
\rowcolor{grpGenRow}
AvatarCLIP~\cite{hong2022avatarclip}      & '22 & Text      & Implicit (NeuS)         & CLIP+SMPL    & \catEM           & \textsf{Gen} \\
Get3DHuman~\cite{xiong2023get3dhuman}     & '23 & Latent    & Tri-plane / SDF         & 3D/2D GAN    & \catEM\tnote{a}  & \cellcolor{grpGenRow}\textsf{Gen} \\
\rowcolor{grpGenRow}
GETAvatar~\cite{zhang2023getavatar}       & '23 & Lat./Pose & Tri-pl.\ + Tex.\ Mesh  & GAN+SMPL     & \catAM           & \textsf{Gen} \\
AvatarCraft~\cite{jiang2023avatarcraft}   & '23 & Text/Img. & NeRF $\to$ Mesh         & Diff.(SDS)   & \catEM           & \cellcolor{grpGenRow}\textsf{Gen} \\
\rowcolor{grpGenRow}
DreamHuman~\cite{kolotouros2023dreamhuman}& '23 & Text      & NeRF + imGHUM           & Diff.(SDS)   & \catEM\tnote{b}  & \textsf{Gen} \\
ChuPa~\cite{kim2023chupa}                & '23 & Text/Img. & SMPL + Displacements    & 2D Diff.     & \catAM           & \cellcolor{grpGenRow}\textsf{Gen} \\
\rowcolor{grpGenRow}
DreamAvatar~\cite{cao2024dreamavatar}     & '24 & Text/Img. & NeRF + SMPL             & Diff.(SDS)   & \catEM\tnote{b}  & \textsf{Gen} \\
TADA!~\cite{liao2024tada}                & '24 & Text      & SMPL-X + Texture        & Diff.\ Prior & \catAM\tnote{d}  & \cellcolor{grpGenRow}\textsf{Gen} \\
%% Feed-Forward / Real-Time
\rowcolor{grpFfbRow}
InstantAvatar~\cite{jiang2023instantavatar}& '23 & Vid.     & Fast Neural Field       & Pose+Masks   & \catRO           & \cellcolor{grpFfbHead}\textcolor{white}{\textsf{FF}} \\
SHERF~\cite{hu2023sherf}                  & '23 & Img./Vid. & NeRF                    & Canon.\ Prior & \catRO          & \cellcolor{grpFfbRow}\textsf{FF} \\
\rowcolor{grpFfbRow}
LRM~\cite{hong2024lrm}                   & '23 & Img.      & Tri-plane / NeRF        & FF-3D        & \catEM\tnote{a}  & \textsf{FF} \\
Human GS~\cite{moreau2024humangs}         & '24 & MV Vid.   & Canonical 3DGS          & Skin.+3D     & \catGS\tnote{c}  & \cellcolor{grpFfbRow}\textsf{FF} \\
\rowcolor{grpFfbRow}
HUGS~\cite{kocabas2024hugs}              & '24 & Img./Vid. & 3DGS + SMPL             & SMPL Prior   & \catGS\tnote{c}  & \textsf{FF} \\
3DGS-Avatar~\cite{qian20243dgsavatar}     & '24 & Vid.      & Deformable 3DGS         & Pose+Masks   & \catGS\tnote{c}  & \cellcolor{grpFfbRow}\textsf{FF} \\
\rowcolor{grpFfbRow}
LHM~\cite{qiu2025lhm}                    & '25 & Img.      & 3DGS                    & 2D/3D+Tr.    & \catGS\tnote{c}  & \textsf{FF} \\
OmniAvatar~\cite{gan2025omniavatar}       & '25 & Audio+Ref & Video Diffusion         & DiT/Video    & \catVD           & \cellcolor{grpFfbRow}\textsf{FF} \\
\bottomrule
\end{tabularx}
\smallskip
\begin{tablenotes}[flushleft]\fontsize{5.5}{6.5}\selectfont
\item[] \textit{Del.:}\; \catAM\,Anim.\ Mesh;\; \catGS\,Anim.\ 3DGS;\; \catEM\,Editable Mesh;\; \catRO\,Render-Only;\; \catVD\,Video.
\item[] \textit{Group:}\;
\textsf{Param}\,=\,Parametric Template;\;
\textsf{Impl}\,=\,Implicit/Hybrid;\;
\textsf{Gen}\,=\,GAN/Diffusion;\;
\textsf{FF}\,=\,Feed-Forward.
\item[$^a$] Requires retopology/mesh extraction.
\item[$^b$] Baked illumination; unsuitable for dynamic relighting.
\item[$^c$] Requires 3DGS engine plugin.
\item[$^d$] Production-ready; LBS-compatible.
\end{tablenotes}
\end{threeparttable}
\end{table*}

\paragraph{From Parametric Templates to Implicit Fields.}\quad
The earliest approaches stayed close to parametric models such as SMPL~\cite{loper2015smpl}: Tex2Shape~\cite{alldieck2019tex2shape} predicted UV-space displacements, CAPE~\cite{ma2020learning} learned pose-dependent clothing offsets, and HybrIK~\cite{li2021hybrik} improved the scaffold itself. These methods offer immediate deployment readiness via standard LBS but cannot represent silhouettes requiring their own topology. Implicit representations---PIFu~\cite{saito2019pifu} and subsequent SMPL-guided variants~\cite{saito2020pifuhd,zheng2021pamir, huang2020arch, xiu2022icon, xiu2023econ, corona2021smplicit, chen2022gdna, corona2023structured3d}---recovered high-resolution clothed geometry without a fixed mesh template (see Table~\ref{tab:full_body_generation}), though occupancy or SDF outputs require mesh extraction and retopology before engine import~\cite{liu2025quadgpt}.

\paragraph{Generative Synthesis (GANs \& Diffusion).}\quad
Once reconstruction pipelines became reliable, the emphasis shifted to synthesizing new identities from images or text. Early GAN-based models~\cite{grigorev2021stylepeople, zhang2022avatargen, xiong2023get3dhuman, zhang2023getavatar} showed that learned priors can produce plausible clothed humans, and score distillation sampling~\cite{poole2023dreamfusion} connected generation to large 2D diffusion priors~\cite{hong2022avatarclip, kolotouros2023dreamhuman, jiang2023avatarcraft, cao2024dreamavatar, han2023headsculpt}. A persistent bottleneck is completing invisible regions with view-consistent appearance: Morphable Diffusion~\cite{chen2024morphable} conditions on an articulated 3D prior for novel-view coherence, and SiTH~\cite{ho2024sith} decomposes back-view hallucination from mesh-guided recovery. ChuPa~\cite{kim2023chupa} and TADA!~\cite{liao2024tada} move closer to production by anchoring geometry and texture to SMPL-X, but most optimization-heavy diffusion methods remain slow and entangle material with illumination.

\paragraph{Feed-Forward LRMs \& Real-Time Representations.}\quad
The most recent direction extends the feed-forward paradigm of LRM~\cite{hong2024lrm} to human avatars. LHM~\cite{qiu2025lhm} regresses dense Gaussians bound to SMPL, yielding animatable avatars in seconds, while deformable 3DGS methods~\cite{jiang2023instantavatar, moreau2024humangs, kocabas2024hugs, qian20243dgsavatar} bind Gaussians to skinning weights for real-time playback. Disco4D~\cite{pang2025disco4d} disentangles body from clothing Gaussians---important where garments must be swapped independently. OmniAvatar~\cite{gan2025omniavatar} and HunyuanVideo-Avatar~\cite{chen2025hunyuanvideo} show that video diffusion can produce temporally coherent performances, though outputs remain closer to generated content than reusable 3D assets.

%% Table 4: Head and Face Synthesis Methods
\begin{table*}[t]
\centering
\caption{Head and face synthesis methods. Methods are color-coded by group; see legend below.}
\label{tab:head_face_synthesis}
\fontsize{6}{7.2}\selectfont
\begin{threeparttable}
\setlength{\tabcolsep}{2pt}
\renewcommand{\arraystretch}{0.95}
\begin{tabularx}{\linewidth}{@{} >{\raggedright\arraybackslash}p{0.16\linewidth}
  >{\centering\arraybackslash}p{0.03\linewidth}
  >{\raggedright\arraybackslash}p{0.09\linewidth}
  >{\raggedright\arraybackslash}X
  >{\raggedright\arraybackslash}p{0.10\linewidth}
  >{\centering\arraybackslash}p{0.045\linewidth}
  >{\centering\arraybackslash}p{0.055\linewidth} @{}}
\toprule
\rowcolor{colHead}
\textcolor{white}{\textbf{Method}} &
\textcolor{white}{\textbf{Yr}} &
\textcolor{white}{\textbf{Input}} &
\textcolor{white}{\textbf{3D Representation}} &
\textcolor{white}{\textbf{Prior}} &
\textcolor{white}{\textbf{Del.}} &
\textcolor{white}{\textbf{Group}} \\
\midrule
%% Morphable + Implicit Neural
\rowcolor{grpMorphRow}
i3DMM~\cite{yenamandra2021i3dmm}          & '21 & MV imgs.  & Implicit SDF (head)     & 3DMM         & \catAN\tnote{a}  & \cellcolor{grpMorphHead}\textcolor{white}{\textsf{Morph}} \\
NerFace~\cite{gafni2021nerface}           & '21 & Mono.\ vid & NeRF+deform.           & 3DMM         & \catAN\tnote{a}  & \cellcolor{grpMorphRow}\textsf{Morph} \\
\rowcolor{grpMorphRow}
EG3D~\cite{chan2022efficient}             & '22 & Lat./img. & Tri-plane NeRF          & None         & \catGN\tnote{d}  & \textsf{Morph} \\
NPHM~\cite{giebenhain2023learning}        & '23 & MV imgs.  & Dual-SDF (id.+expr.)    & 3DMM         & \catAN\tnote{a}  & \cellcolor{grpMorphRow}\textsf{Morph} \\
\rowcolor{grpMorphRow}
Next3D~\cite{sun2023next3d}              & '23 & Pose+lat. & Tri-pl.+tex.\ rast.    & FLAME        & \catGN           & \textsf{Morph} \\
PanoHead~\cite{an2023panohead}            & '23 & Lat./img. & Depth-aware tri-grid    & None         & \catGN\tnote{d}  & \cellcolor{grpMorphRow}\textsf{Morph} \\
\rowcolor{grpMorphRow}
RODIN~\cite{wang2023rodin}               & '23 & Text/img. & Diffusion tri-plane     & None         & \catGN\tnote{b}  & \textsf{Morph} \\
HeadSculpt~\cite{han2023headsculpt}       & '23 & Text      & SDS NeRF                & SMPLX        & \catGN\tnote{b}  & \cellcolor{grpMorphRow}\textsf{Morph} \\
%% Mesh-Anchored Gaussians
\rowcolor{grpGsRow}
GaussianAvatars~\cite{qian2024gaussianavatars} & '24 & MV vid. & 3DGS on FLAME tris   & FLAME        & \catGS           & \cellcolor{grpGsHead}\textcolor{white}{\textsf{GS}} \\
FlashAvatar~\cite{xiang2024flashavatar}   & '24 & Mono.\ vid & 3DGS+FLAME embed.     & FLAME        & \catGS\tnote{c}  & \cellcolor{grpGsRow}\textsf{GS} \\
\rowcolor{grpGsRow}
MonoGaussAvt~\cite{chen2024monogaussianavatar} & '24 & Mono.\ vid & 3DGS+FLAME        & FLAME        & \catGS           & \textsf{GS} \\
RGCA~\cite{saito2024relightable}          & '24 & MV vid.   & PBR 3DGS                & FLAME        & \catGS\tnote{e}  & \cellcolor{grpGsRow}\textsf{GS} \\
%% Feed-Forward Reconstruction
\rowcolor{grpFfrecRow}
GAGAvatar~\cite{chu2024generalizable}     & '24 & 1 img.    & Generalizable 3DGS      & FLAME        & \catGS           & \cellcolor{grpFfrecHead}\textcolor{white}{\textsf{FFrec}} \\
Arc2Avatar~\cite{gerogiannis2025arc2avatar}& '25 & 1 img.    & 3DGS (ID-guided)       & FLAME        & \catGS           & \cellcolor{grpFfrecRow}\textsf{FFrec} \\
\rowcolor{grpFfrecRow}
HRAvatar~\cite{zhang2025hravatar}         & '25 & Mono.\ vid & PBR 3DGS (rough.+Fres.)& FLAME       & \catGS\tnote{e}  & \textsf{FFrec} \\
LAM~\cite{he2025lam}                      & '25 & Sparse    & Canonical 3DGS          & FLAME        & \catGS\tnote{c}  & \cellcolor{grpFfrecRow}\textsf{FFrec} \\
\rowcolor{grpFfrecRow}
Avat3r~\cite{kirschstein2025avat3r}       & '25 & Sparse    & Canonical 3DGS          & FLAME        & \catGS           & \textsf{FFrec} \\
%% Rendering and Animation
\rowcolor{grpAnimRow}
SadTalker~\cite{zhang2023sadtalker}       & '23 & Audio+img. & 2D render+3D coeff.    & FLAME        & \catVD\tnote{f}  & \cellcolor{grpAnimHead}\textcolor{white}{\textsf{Anim}} \\
TexTalker~\cite{li2025towards}            & '25 & Audio+img. & 3D mesh+diff.\ tex.    & FLAME        & \catGS\tnote{f}  & \cellcolor{grpAnimRow}\textsf{Anim} \\
\bottomrule
\end{tabularx}
\smallskip
\begin{tablenotes}[flushleft]\fontsize{5.5}{6.5}\selectfont
\item[] \textit{Del.:}\; \catAN\,Anim.\ Neural;\; \catGN\,Novel-ID Gen.;\; \catGS\,Anim.\ 3DGS;\; \catVD\,Video.
\item[] \textit{Group:}\;
\textsf{Morph}\,=\,Morphable+Neural;\;
\textsf{GS}\,=\,Mesh-Anchored GS;\;
\textsf{FFrec}\,=\,FF Recon.;\;
\textsf{Anim}\,=\,Rendering/Anim.
\item[$^a$] Implicit output; requires mesh extraction before engine import.
\item[$^b$] Baked illumination; unsuitable for dynamic relighting.
\item[$^c$] Production-ready; LBS-compatible (FlashAvatar $>$300\,FPS, LAM $>$280\,FPS).
\item[$^d$] Novel-identity generation; expression control limited to camera pose.
\item[$^e$] Full PBR decomposition (albedo, specular, roughness/Fresnel).
\item[$^f$] Speech-driven animation.
\end{tablenotes}
\end{threeparttable}
\end{table*}

\subsection{Head and Face Synthesis}
\label{sec:head_face_synthesis}

The head and face demand considerably higher fidelity than the body due to human sensitivity to facial micro-expressions and gaze direction. A production-ready head asset must exhibit stable identity across viewpoints, provide explicit animation mechanisms (e.g., blendshapes), and support physically based relighting. Work in this area spans morphable-model-based implicit generation, real-time Gaussian representations bound to mesh trackers, and feed-forward reconstruction; Table~\ref{tab:head_face_synthesis} provides a structured comparison.

\paragraph{Morphable Model Priors and Implicit Neural Fields.}\quad
Traditional workflows relied on 3DMM or the FLAME model~\cite{li2017flame} to separate shape, pose, and expression into independent subspaces mapping to engine bone hierarchies. To overcome the fixed topology limitation, i3DMM~\cite{yenamandra2021i3dmm} and NPHM~\cite{giebenhain2023learning} formulated the entire head as a continuous SDF, and NerFace~\cite{gafni2021nerface} extended NeRF with expression-conditioned deformation. Generative efforts used tri-planes (EG3D~\cite{chan2022efficient}, Next3D~\cite{sun2023next3d}, PanoHead~\cite{an2023panohead}) and score distillation~\cite{wang2023rodin, han2023headsculpt} for novel-identity generation (Table~\ref{tab:head_face_synthesis}, Morph group). Despite their visual quality, all implicit and volumetric approaches require ray-marching, conflicting with rasterization-based engines.

\paragraph{Mesh-Anchored 3D Gaussians (Per-Scene Optimization).}\quad
Research from 2024 onward shifted toward anchoring 3D Gaussians to explicit parametric mesh trackers as a way to reconcile generative quality with real-time engine performance. By binding Gaussians to the triangles of a FLAME mesh, GaussianAvatars~\cite{qian2024gaussianavatars}, FlashAvatar~\cite{xiang2024flashavatar}, and MonoGaussianAvatar~\cite{chen2024monogaussianavatar} combine the topological flexibility of point-based rendering with the animation control of standard blendshapes. Because each Gaussian inherits the linear blend skinning weights of FLAME, the avatar responds to the same blendshape coefficients already expected by engine animation graphs, removing the need for any representation conversion at deployment time. FlashAvatar achieves rendering speeds exceeding 300 FPS at high resolutions, satisfying the real-time requirements of interactive games.

\paragraph{Generalizable Feed-Forward Reconstruction.}\quad
Per-subject Gaussian methods initially required hours of multi-view video optimization. Feed-forward approaches have largely resolved this bottleneck. These models are typically pre-trained on large-scale multi-view datasets such as NeRSemble~\cite{kirschstein2023nersemble}, which provides 220 subjects captured across 16 synchronized cameras at high frame rates, a scale that enables genuine cross-identity generalization rather than per-scene memorization. GAGAvatar~\cite{chu2024generalizable} and Arc2Avatar~\cite{gerogiannis2025arc2avatar} infer Gaussian parameters directly from a single image using facial foundation models to maintain identity fidelity without test-time optimization. LAM~\cite{he2025lam} and Avat3r~\cite{kirschstein2025avat3r} process sparse, unconstrained image inputs through Vision Transformers to output animatable Gaussians, with LAM achieving 280 FPS on desktop GPUs and 35 FPS on mobile devices. Because predictions reside in a canonical space aligned with standard linear blend skinning weights, the outputs integrate directly with engine animation graphs.

\paragraph{Rendering and Animation.}\quad
A face with baked-in illumination appears inconsistent under dynamic game lighting. Relightable Gaussian Codec Avatars~\cite{saito2024relightable} decompose facial appearance into albedo, diffuse, and specular components, and HRAvatar~\cite{zhang2025hravatar} extends relightable generation to monocular inputs with explicit roughness and Fresnel modeling. For speech-driven animation, TexTalker~\cite{li2025towards} synthesizes audio-synchronized wrinkle maps alongside geometric deformation, while SadTalker~\cite{zhang2023sadtalker} maps acoustic features to FLAME motion coefficients to produce talking-head video from a single image. Though SadTalker does not output a 3D asset, its decoupled audio-to-motion formulation is widely adopted as a driving module within full 3D pipelines.

\subsection{The Production-Ready Bridge: Automatic Rigging and Skinning}
\label{sec:rigging_skinning}

Even structurally flawless geometry remains a static sculpture until equipped with a skeleton and skinning weights, making the bridge to engine-compliant articulated assets the ultimate hurdle in the 3D pipeline. Three paradigms have emerged.

\paragraph{Parametric Weight Inheritance.} \quad
Frameworks such as TADA!~\cite{liao2024tada} and ChuPa~\cite{kim2023chupa} sculpt details as displacements on a canonical SMPL-X template, natively inheriting its skeletal hierarchy and LBS weights for immediate animation compatibility. The limitation is that projecting body skinning weights onto loose garments (long skirts, coats) tends to produce candy-wrapper artifacts during large joint rotations.

\paragraph{Topology-Agnostic Neural Prediction.} \quad
To support arbitrary morphologies (monsters, stylized creatures, armored avatars), RigNet~\cite{xu2020rignet} formulated skeleton extraction and skinning weight prediction as an end-to-end GNN task on arbitrary meshes. SkinningNet~\cite{mosella2022skinningnet} introduced a two-stream GNN with multi-aggregator convolutions for heterogeneous skeletal topologies, and DeePSD~\cite{bertiche2021deepsd} extended this to garments of arbitrary topology by mapping clothing meshes directly to engine-compatible skinning weights and blend shapes via unsupervised physics losses, explicitly resolving body--garment collisions without topology constraints. These graph-based methods significantly reduce manual labor for animating diverse AI-generated assets.

\paragraph{Explicit Kinematic Binding for Neural Representations.} The point-based nature of 3DGS conflicts with polygon-based LBS pipelines of game engines. Recent Large Avatar Models resolve this by integrating LBS directly into the generative formulation: LAM~\cite{he2025lam} and HRAvatar~\cite{zhang2025hravatar} map Gaussian attributes to a canonical space governed by learnable LBS matrices and blendshapes, achieving $>$200 FPS on standard GPUs while remaining drivable by standard engine animation graphs.

\begin{figure*}[t]
    \centering
    \includegraphics[width=\textwidth]{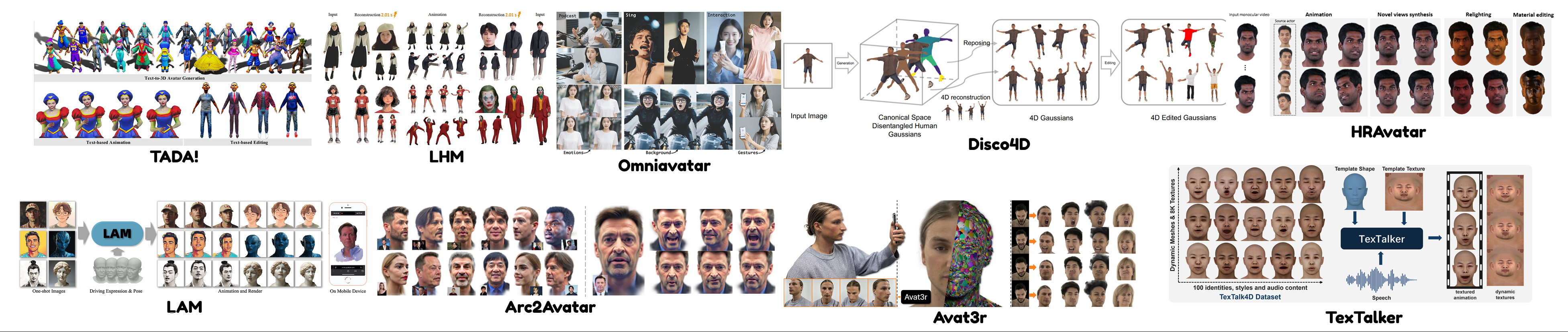}
    \caption{Character and avatar generation methods (2024--2025). Full-body synthesis: TADA!~\cite{liao2024tada}, LHM~\cite{qiu2025lhm}, OmniAvatar~\cite{gan2025omniavatar}, Disco4D~\cite{pang2025disco4d}; Head and face generation: HRAvatar~\cite{zhang2025hravatar}, LAM~\cite{he2025lam}, Arc2Avatar~\cite{gerogiannis2025arc2avatar}, Avat3r~\cite{kirschstein2025avat3r}; Speech-driven animation: TexTalker~\cite{li2025towards}.}
    \label{fig:character_avatar_methods}
\end{figure*}

\section{Scenes and Environments}
\label{sec:scenes_environments}
Scenes and environments differ from standalone props or characters: a scene is a coordinated arrangement of assets, free space, lighting, materials, and interaction states that jointly determine navigability, affordance perception, and player actions. A production-ready scene generator must therefore preserve traversability, collision validity, asset grounding, and compatibility with engine systems such as physics, scripting, and level editing.

Recent progress has shifted emphasis from static indoor arrangement toward controllable, executable world construction, combining procedural priors, diffusion or transformer layout models, and LLM-based planners to produce structured layouts, retrieve concrete assets, and export to engines~\cite{yang2024holodeck,hu2024scenecraft,tang2025unrealllm}. Having consolidated major scene corpora in Section~\ref{sec:data_scenes}, the focus here is on how generation methods translate heterogeneous data priors into playable environments.

We therefore organize this section according to the lifecycle of a playable environment. Layout generation first establishes the structural scaffold of rooms, streets, or open terrain. Scene population and asset grounding then translate abstract slots and relations into concrete reusable content. The discussion next expands to terrain, skyboxes, and world-scale continuity, where memory efficiency and visual coherence become central. Finally, we close with the emerging direction of playable and interactive world models, in which scenes are evaluated not only by appearance but also by actionability, controllability, and engine integration. Table~\ref{tab:scene_generation} summarizes representative methods from this production-ready perspective.

\providecommand{\scenebadge}[2]{\begingroup\setlength{\fboxsep}{1pt}\colorbox{#1}{\scriptsize\textsf{#2}}\endgroup}
\providecommand{\stageLay}{\scenebadge{blue!12}{Lay}}
\providecommand{\stageGround}{\scenebadge{green!12}{Gnd}}
\providecommand{\stageExec}{\scenebadge{orange!18}{Exe}}
\providecommand{\stageWorld}{\scenebadge{teal!16}{Wld}}
\providecommand{\stagePlay}{\scenebadge{red!12}{Play}}
\providecommand{\tagProc}{\scenebadge{gray!22}{Proc.}}
\providecommand{\tagLang}{\scenebadge{blue!18}{Lang.}}
\providecommand{\tagRet}{\scenebadge{green!18}{Ret.}}
\providecommand{\tagAff}{\scenebadge{orange!22}{Aff.}}
\providecommand{\tagPhys}{\scenebadge{red!14}{Phys.}}
\providecommand{\tagEdit}{\scenebadge{yellow!22}{Edit}}
\providecommand{\tagEng}{\scenebadge{teal!18}{Engine}}
\providecommand{\tagScale}{\scenebadge{violet!16}{Scale}}

\begin{table*}[t]
\centering
\caption{Representative scene and environment generation methods viewed by pipeline stage.}
\label{tab:scene_generation}
\fontsize{6}{7.2}\selectfont
\begin{threeparttable}
\setlength{\tabcolsep}{2pt}
\renewcommand{\arraystretch}{0.95}
\begin{tabularx}{\linewidth}{@{} >{\raggedright\arraybackslash}p{0.14\linewidth}
  >{\centering\arraybackslash}p{0.03\linewidth}
  >{\raggedright\arraybackslash}p{0.24\linewidth}
  >{\raggedright\arraybackslash}p{0.24\linewidth}
  >{\raggedright\arraybackslash}X
  >{\centering\arraybackslash}p{0.05\linewidth} @{}}
\toprule
\rowcolor{colHead}
\textcolor{white}{\textbf{Method}} &
\textcolor{white}{\textbf{Yr}} &
\textcolor{white}{\textbf{Main Driver}} &
\textcolor{white}{\textbf{Output Form}} &
\textcolor{white}{\textbf{Game-Oriented Tags}} &
\textcolor{white}{\textbf{Group}} \\
\midrule
%% Layout Planning
\rowcolor{grpLayRow}
ATISS~\cite{paschalidou2021atiss}          & '21 & object-set transformer     & room layouts              & \tagEdit              & \cellcolor{grpLayHead}\textcolor{white}{\textsf{Lay}} \\
ProcTHOR~\cite{deitke2022procthor}         & '22 & procedural rules           & interactive houses        & \tagProc~\tagPhys~\tagScale & \cellcolor{grpLayRow}\textsf{Lay} \\
\rowcolor{grpLayRow}
Pose2Room~\cite{nie2022pose2room}          & '22 & human activities            & affordance-aware layouts  & \tagAff~\tagPhys      & \textsf{Lay} \\
DiffuScene~\cite{tang2024diffuscene}       & '24 & diffusion + retrieval       & furnished indoor scenes   & \tagRet~\tagEdit      & \cellcolor{grpLayRow}\textsf{Lay} \\
\rowcolor{grpLayRow}
Holodeck~\cite{yang2024holodeck}           & '24 & LLM planning               & embodied environments     & \tagLang~\tagRet~\tagPhys & \textsf{Lay} \\
%% Grounding and Construction
\rowcolor{grpGndRow}
AnyHome~\cite{fu2024anyhome}              & '24 & structured text-to-home     & house-scale scenes        & \tagLang~\tagRet~\tagEdit & \cellcolor{grpGndHead}\textcolor{white}{\textsf{Gnd}} \\
Open-Universe~\cite{aguinakang2024openuniverse} & '24 & LLM programs + solver  & retrieved asset scenes    & \tagLang~\tagRet~\tagEdit & \cellcolor{grpGndRow}\textsf{Gnd} \\
\rowcolor{grpGndRow}
SceneCraft~\cite{hu2024scenecraft}         & '24 & Blender code agent          & executable scene scripts  & \tagLang~\tagEng~\tagEdit & \textsf{Gnd} \\
PhyScene~\cite{yang2024physcene}           & '24 & physics-guided diffusion    & interactable scenes       & \tagAff~\tagPhys      & \cellcolor{grpGndRow}\textsf{Gnd} \\
\rowcolor{grpGndRow}
UnrealLLM~\cite{tang2025unrealllm}         & '25 & Unreal PCG agents           & engine scenes             & \tagLang~\tagEng~\tagPhys & \textsf{Gnd} \\
Layout2Scene~\cite{chen2025layout2scene}   & '25 & layout-guided diffusion     & holistic scenes           & \tagEdit              & \cellcolor{grpGndRow}\textsf{Gnd} \\
%% World-Scale
\rowcolor{grpWldRow}
Text2Light~\cite{chen2022text2light}       & '22 & text-conditioned HDR        & skybox and lighting       & \tagScale             & \cellcolor{grpWldHead}\textcolor{white}{\textsf{Wld}} \\
Infinigen In.~\cite{raistrick2024infinigenindoors} & '24 & photorealistic procedural & indoor worlds          & \tagProc~\tagScale~\tagEng & \cellcolor{grpWldRow}\textsf{Wld} \\
\rowcolor{grpWldRow}
WorldCraft~\cite{liu2025worldcraft}        & '25 & agentic world editing       & customized worlds         & \tagLang~\tagScale~\tagEdit & \textsf{Wld} \\
LayerPano3D~\cite{shuaiyang2025layerpano3d}     & '25 & layered panorama synthesis  & explorable panoramas      & \tagScale~\tagEdit    & \cellcolor{grpWldRow}\textsf{Wld} \\
\bottomrule
\end{tabularx}
\smallskip
\begin{tablenotes}[flushleft]\fontsize{5.5}{6.5}\selectfont
\item[] \textit{Tags:}\;
\tagProc~procedural;\;
\tagLang~language;\;
\tagRet~retrieval;\;
\tagAff~human/affordance;\;
\tagPhys~physics;\;
\tagEdit~editable;\;
\tagEng~engine-native;\;
\tagScale~world-scale.\quad
\textit{Group:}\;
\textsf{Lay}\,=\,Layout Planning;\;
\textsf{Gnd}\,=\,Grounding \& Construction;\;
\textsf{Wld}\,=\,World-Scale.
\end{tablenotes}
\end{threeparttable}
\end{table*}

\subsection{Layout Generation as the Structural Foundation}
\label{sec:scene_layout}

Layout generation defines the structural validity of a scene before any concrete geometry is instantiated, determining whether rooms preserve traversable free space, streets admit later asset insertion, and camera paths or embodied agents can move without immediate collision. As summarized in Section~\ref{sec:data_scenes}, the relevant priors now span annotated indoor corpora such as 3D-FRONT and Structured3D as well as procedural worlds such as ProcTHOR and Infinigen Indoors~\cite{fu20213d,zheng2020structured3d,deitke2022procthor,raistrick2024infinigenindoors}.

\paragraph{Constraint-driven and procedural layout generation.}\quad
ProcTHOR procedurally generates interactive houses supporting navigation and manipulation benchmarks~\cite{deitke2022procthor}. Infinigen Indoors extends this toward photorealistic indoor spaces with export paths to real-time simulators~\cite{raistrick2024infinigenindoors}. These systems demonstrate a practical design choice: when the scaffold is produced by explicit rules, collision checks and room connectivity can be enforced before expensive synthesis begins.

\paragraph{Learned layout priors and relation models.}\quad
ATISS showed that autoregressive transformers over unordered object sets can synthesize plausible room layouts conditioned on floor plans~\cite{paschalidou2021atiss}. Subsequent work incorporated stronger functional priors: Pose2Room and MIME use human motion for contact-compatible placement~\cite{nie2022pose2room,yi2023mime}, DiffuScene models composition via denoising over joint object attributes~\cite{tang2024diffuscene}, and PhyScene adds explicit physical and reachability guidance as part of the layout objective~\cite{yang2024physcene}.

\paragraph{Language-guided layout planning.}\quad
Recent systems treat natural language as a front-end authoring interface while relying on structured intermediate forms for placement. LayoutGPT~\cite{feng2023layoutgpt} uses an LLM to produce indoor layouts from text, Holodeck~\cite{yang2024holodeck} combines language planning with asset retrieval and spatial optimization for embodied environments, LLplace~\cite{yang2024llplace} supports dialogue-driven editing, and CityCraft~\cite{deng2024citycraft} extends language-guided planning to city-scale generation. The common trend is clear: language specifies intent, but reliable layout still depends on checkable representations such as floor plans or optimization variables.

\subsection{The Production-Ready Bridge: Scene Population and Asset Grounding}
\label{sec:scene_population}

After layout planning, a scene still exists only as a scaffold of slots, relations, and free space. A deployable environment requires concrete assets, materials, states, and scripts. This bridge from abstract layout to reusable content is central to game pipelines, because most studios do not synthesize every mesh from scratch. Instead, they populate scenes through libraries of curated or retrieved assets whose semantics, collision meshes, and material channels remain editable after placement.

\paragraph{Retrieval-based and open-vocabulary grounding.}\quad
Retrieval-centered systems preserve asset separability---a chair remains replaceable rather than a fused region in a scene-wide neural field. AnyHome translates open-vocabulary text into house-scale structured scenes with explicit hierarchies~\cite{fu2024anyhome}. Open-Universe Indoor Scene Generation couples LLM-generated layout programs with uncurated object databases~\cite{aguinakang2024openuniverse}. Holodeck~\cite{yang2024holodeck} follows a similar retrieve-then-populate philosophy in embodied environments.

\paragraph{Script- and agent-driven scene construction.}\quad
A second trend is executable representations. SceneCraft converts text into Blender-executable Python code with visual feedback for spatial error repair~\cite{hu2024scenecraft}. 3D-GPT generalizes procedural modeling instructions as controllable intermediates~\cite{sun20253d}. UnrealLLM maps language to Unreal Engine blueprints and spline-based control~\cite{tang2025unrealllm}. Executable descriptions are slower to design than direct pixels but far easier to audit, edit, version, and integrate into engine toolchains.

\paragraph{Human- and physics-aware population.}\quad
Placement must respect agent motion envelopes and contact patterns. Pose2Room and MIME show that activity priors improve object functionality~\cite{nie2022pose2room,yi2023mime}, PhyScene makes articulation and collision explicit during synthesis~\cite{yang2024physcene}, and PhysGen3D extends the discussion to miniature interactive worlds combining generation with physics simulation~\cite{chen2025physgen3d}.

\paragraph{Joint layout-and-content synthesis.}\quad
Text2Room extracts textured scene meshes from 2D diffusion priors~\cite{hollein2023text2room}. Layout2Scene conditions diffusion on semantic layouts for improved controllability~\cite{chen2025layout2scene}. These systems are promising when they preserve intermediate structure but harder to deploy when geometry, texture, and identity are optimized into a single inseparable representation.

\subsection{Terrain, Skybox, and World-Scale Environment Generation}
\label{sec:world_scale}

The scene literature often concentrates on furnished rooms, yet many games derive their identity from outdoor terrain, distant backdrops, and large continuous worlds. Once the target expands from a room to a district or explorable landscape, local plausibility remains necessary but no longer sufficient. The generator must also maintain continuity across long traversal distances, manage memory and level-of-detail, and support lighting or atmosphere that remains stable under free camera motion.

\paragraph{Procedural terrain and large-scale world scaffolds.}\quad
The graphics community has long relied on procedural generation for terrain, vegetation, and ecosystem diversity. Infinigen shows how large outdoor environments can be synthesized with strong appearance diversity and geographic extent~\cite{raistrick2023infinite}. Infinigen Indoors brings a related philosophy to interior spaces, with export-oriented processing for materials, collision, and simulator compatibility~\cite{raistrick2024infinigenindoors}. More recent LLM-driven systems such as CityCraft and WorldCraft suggest a hybrid future in which procedural world scaffolds remain the backbone for scale, while language agents control composition, customization, and iterative revision~\cite{deng2024citycraft,liu2025worldcraft}. This hybrid design is attractive for games because hand-authored rules still provide stable controllability at scale, whereas learned modules improve accessibility and variation.

\paragraph{Neural outdoor and city-scale generation.}\quad
Procedural scaffolds offer deterministic control but require extensive rule engineering for each biome or urban archetype.
Learning-based approaches complement this by capturing appearance distributions directly from real-world data.
CityDreamer~\cite{xie2024citydreamer} addresses unbounded urban synthesis by factoring city scenes into a stuff field encoding large-scale ground structure and object-class-specific thing fields for individual buildings, enabling kilometre-scale traversal without tiling seams.
For terrain synthesis, neural height-map and texture generation methods now match classical noise-based generators in visual diversity while remaining compatible with standard engine height-field importers.
A persistent challenge at open-world scale is \emph{ecological coherence}: biome transitions, road-network topology, and landmark distribution follow geographic constraints that neither pure noise models nor image-domain diffusion reliably capture.
Hybrid strategies---procedural macro-structure with neural micro-variation---remain the most production-viable path for games that must balance kilometre-scale extent with per-region ecological plausibility.

\paragraph{Panoramic, skybox, and immersive environment generation.}\quad
Text2Light synthesizes HDR panoramas from text, providing controllable skybox and environment-light creation~\cite{chen2022text2light}. LayerPano3D decomposes panoramas into layered depth-aware structures liftable into explorable 3D Gaussian scenes~\cite{shuaiyang2025layerpano3d}. These works bridge the gap between static backdrops and traversable environments, though editability and deep interaction remain limited.

\paragraph{Holistic world assembly and cross-scale control.}\quad
WorldCraft uses LLM agents to coordinate object customization, layout optimization, and trajectory control~\cite{liu2025worldcraft}. CityCraft applies similar logic to urban generation across larger spatial footprints~\cite{deng2024citycraft}. Layout2Scene shows that explicit semantic structure remains useful even in joint synthesis~\cite{chen2025layout2scene}. Together, these systems suggest future game environments will be assembled from a mixture of procedural terrain, explicit asset instances, panoramic background layers, and executable control scripts rather than from one monolithic generator.

\subsection{Toward Playable and Interactive World Models}
\label{sec:playable_world_models}

The preceding sections address scene \emph{appearance}: how plausible layouts are generated, how concrete assets are grounded, and how large-scale continuity is maintained.
Playability introduces a qualitatively different requirement.
A game-ready world must support navigation, collision handling, object state transitions, script binding, and iterative re-authoring.
Distributional metrics such as FID or CLIP similarity cannot measure whether a generated doorway is wide enough for an agent to traverse, whether a placed switch can be toggled, or whether the resulting scene can be imported into an engine without error.
Closing this gap requires treating interactivity, physical correctness, and executability as first-class generation objectives rather than post-hoc validation criteria.

\paragraph{Executable scene representations.}\quad
The most direct response to the playability requirement is to generate scenes in a form that engines can natively interpret.
SceneCraft converts natural language specifications into Blender-executable Python scripts, using a visual feedback loop to detect and repair spatial inconsistencies before finalization~\cite{hu2024scenecraft}.
UnrealLLM maps language to Unreal Engine Procedural Content Generation nodes and spline blueprints, yielding scenes that are immediately editable within the engine's existing toolchain~\cite{tang2025unrealllm}.
3D-GPT frames procedural modeling instructions as language-conditioned programs that govern object instantiation, parameter sweeps, and layout constraints~\cite{sun20253d}.
These systems share an architectural logic with direct implications for game production: executable outputs are \emph{auditable} (scripts can be reviewed and diffed), \emph{version-controllable} (they integrate with standard development workflows), and \emph{iterable} (the language model can revise a script rather than regenerate from scratch).
The primary limitation is spatial precision: layout errors arising from imprecise code generation require either rule-based geometric checkers or round-trip visual feedback to detect and correct.

\paragraph{Physics-grounded interactivity.}\quad
Beyond collision avoidance, interactive game scenes require that objects rest stably, respond predictably to simulation, and can be activated or manipulated by agents.
PhyScene integrates physical feasibility objectives directly into the layout diffusion process, penalizing interpenetration and verifying object stability through short forward-simulation rollouts after each generation step~\cite{yang2024physcene}.
PhysGen3D extends this to miniature interactive worlds in which object affordances and dynamic behaviors are modeled as explicit generation targets~\cite{chen2025physgen3d}.
Holodeck constructs embodied environments whose asset placements are validated against navigation and manipulation benchmarks, treating reachability as a layout constraint rather than a post-hoc check~\cite{yang2024holodeck}.
These systems demonstrate that physical constraints can enter the generation loop rather than serving only as a filtering stage.
The remaining challenge is scope: current methods address \emph{placement-level} physics---whether objects are stable and agents can move---while \emph{intrinsic} physical properties such as mass, friction coefficients, and articulation limits must still be assigned through separate annotation pipelines or manual authoring.

\paragraph{Evaluation gaps and architectural outlook.}\quad
The field still lacks evaluation protocols commensurate with game-production requirements.
Standard appearance metrics are silent on traversability, interaction coverage, engine import success, and revision fidelity.
ProcTHOR provides a partial exception by coupling generation with navigation and manipulation benchmarks~\cite{deitke2022procthor}, and PhyScene measures post-simulation stability, but neither targets the full engine import pipeline.
Looking forward, the most production-viable architecture is not a single universal generator but a layered system: procedural scaffolds for stable large-scale layout, LLM-driven executable specifications for intent and iterative revision, retrieval-based grounding for asset identity, and physics-aware objectives for interaction validity.
In this hierarchy, LLMs, diffusion models, and procedural systems are complementary stages within a production-oriented pipeline, each responsible for a different level of the scene hierarchy.

\section{Evaluation Metrics and Protocols}
\label{sec:evaluation}
Evaluation in 3D generation has historically borrowed metrics from image synthesis and point cloud analysis. The production-pipeline perspective of this survey, however, exposes a significant gap: most existing metrics assess visual or geometric fidelity in isolation, while attributes determining whether an asset can be deployed in an interactive engine---topological correctness, UV quality, material separability, rig robustness, and runtime efficiency---remain largely unmeasured. This section therefore surveys the principal evaluation metrics and protocols, organized from geometric fidelity through appearance and asset usability to scene-level quality, and identifies the standardization gaps that limit assessment of production readiness.

\subsection{Geometric and Appearance Fidelity}
\label{sec:eval_fidelity}

At the geometric level, standard per-instance metrics include Chamfer Distance (CD)~\cite{fan2017point}, Earth Mover's Distance (EMD)~\cite{fan2017point}, F-Score at threshold~$\tau$, and Normal Consistency; distribution-level metrics---Coverage (COV), Minimum Matching Distance (MMD), and 1-Nearest Neighbor Accuracy (1-NNA)~\cite{yang2019pointflow,achlioptas2018learning}---assess diversity, fidelity, and separability.

On the appearance side, PSNR, SSIM, and LPIPS~\cite{zhang2018unreasonable} compare rendered views against ground truth, while FID~\cite{heusel2017gans}, KID, CLIP Score, and CLIP R-Precision~\cite{radford2021learning} evaluate distributional realism and semantic alignment. A notable gap concerns PBR material fidelity: production-ready assets require separated albedo, roughness, and metallic channels, yet no widely adopted standardized metric yet exists for material separation quality. Paint3D~\cite{zeng2024paint3d} and MaterialMVP~\cite{he2025materialmvp} have proposed relighting-based consistency tests, but these remain paper-specific protocols.

\subsection{Asset Usability and Deployment Readiness}
\label{sec:eval_usability}

From the production-pipeline perspective, the most critical yet neglected evaluation dimension is whether a generated asset can be used in a downstream engine or tool. This usability assessment spans several axes. UV parameterization quality includes stretch and angular distortion, seam visibility, chart packing efficiency, and overlap detection---metrics that are well established in geometry processing but rarely reported in generative papers. Recent UV-aware methods such as ArtUV~\cite{chen2025artuv} and PartUV~\cite{wang2025partuv} have begun to include chart-level statistics, but a comprehensive UV benchmark against production standards does not yet exist.

For characters and deformable objects, rig quality determines animation readiness. Skinning weight smoothness, self-intersection rates under extreme poses, joint limit violations, and motion retargeting success together characterize skeleton compatibility with existing animation pipelines. RigNet~\cite{xu2020rignet} introduced a basic evaluation protocol, but it covers only a narrow range of skeleton types and does not assess deformation under motion.

Perhaps the most operationally relevant test is engine import success: can the asset be imported into Unreal Engine, Unity, or Blender and rendered at interactive frame rates without manual intervention? Triangle budget compliance, draw call efficiency, and collision mesh quality further determine runtime viability. Holodeck~\cite{yang2024holodeck} and UnrealLLM~\cite{tang2025unrealllm} have demonstrated engine-integrated generation but neither reported systematic usability metrics. Topology quality---manifoldness, watertightness, genus correctness, quad ratio, edge flow alignment---is essential for production but almost never quantified. MeshAnything~\cite{chen2024meshanything} and DeepMesh~\cite{zhao2025deepmesh} report basic topology statistics, yet no standardized topology evaluation suite exists. This disconnect between training objectives (typically rendering losses) and deployment requirements (structural correctness) remains a central obstacle.

\subsection{Scene-Level and Human Evaluation}
\label{sec:eval_scene}

Scene evaluation introduces spatial, physical, and functional constraints beyond single-asset quality. Physical plausibility requires stable resting without interpenetration; PhyScene~\cite{yang2024physcene} measures object displacement after short physics simulation. Traversability, via NavMesh connectivity and navigation success rates~\cite{deitke2022procthor}, determines whether generated environments admit meaningful agent exploration. Functional plausibility---whether chairs are sittable, doors openable, drawers pullable---represents the most demanding criterion, tested by human activity compatibility~\cite{nie2022pose2room,yi2023mime} and physics-based interaction~\cite{chen2025physgen3d}.

Human evaluation remains indispensable for holistic quality assessment. Standard protocols include A/B preference tests, Likert-scale ratings across dimensions (geometry, texture, alignment, realism), and perceptual discrimination studies such as GameNGen's visual Turing test~\cite{valevski2025gamengen}. Professional evaluation by technical artists---judging topology, edge flow, UV layout, material correctness, and estimated cleanup time---would be the most informative signal for production readiness but remains largely absent from the literature.

Among emerging benchmarks, T$^{3}$Bench~\cite{he2024t3bench} provides the first structured text-to-3D evaluation suite with automatic quality and alignment metrics across three difficulty levels. The TRELLIS evaluation protocol~\cite{xiang2025structured, xiang2025trellis2} consolidates geometry and appearance scores for large-scale comparison. Nevertheless, the field still lacks a unified production-readiness benchmark spanning geometric accuracy, topological correctness, material quality, rig robustness, and engine import success. Building such a benchmark requires professionally produced reference assets with full pipeline annotations and automated engine integration tests. Without it, the community will continue optimizing individual metrics that do not jointly guarantee deployment readiness.

\section{Open Challenges and Future Directions}
\label{sec:open_challenges}
The evaluation gaps identified above point toward broader structural challenges the field must address before 3D generation can serve as infrastructure for interactive worlds. This section synthesizes five open research directions, from data and pipeline-level concerns to forward-looking questions about world models, scaling, and societal responsibility.

\subsection{Data, Controllability, and the Assetization Bottleneck}
\label{sec:challenge_pipeline}

Three tightly coupled bottlenecks constrain the practical utility of current 3D generation methods. The first is data. As discussed in Section~\ref{sec:data_foundations}, the largest public repository, Objaverse-XL~\cite{deitke2023objaversexl}, exceeds ten million assets but inherits the quality heterogeneity of its web-crawled sources, while curated datasets such as Google Scanned Objects~\cite{downs2022google} offer production-grade annotations at a scale two orders of magnitude smaller. To our knowledge, no existing dataset simultaneously provides manifold topology, artist-quality UV layouts, complete PBR channels, skeletal rigs, and natural-language descriptions. The character domain is even more fragmented, with body shape, facial identity, motion, and clothing deformation scattered across incompatible datasets. Procedural generation~\cite{raistrick2023infinite,raistrick2024infinigenindoors}, automated quality filtering~\cite{deitke2023objaverse,luo2023cap3d}, and annotation harvesting from game engines are promising mitigation strategies, but bridging the scale--quality gap remains a prerequisite for next-generation generative models.

The second bottleneck is controllability. Current methods overwhelmingly operate in a one-shot paradigm---the user provides a prompt and receives a complete asset with no opportunity for iterative refinement. This conflicts with typical professional workflows, in which artists revise assets through many editing cycles. Part-aware methods such as OmniPart~\cite{yang2025omnipart} and PartCrafter~\cite{lin2025partcrafter} enable component-level editing, and SceneCraft~\cite{hu2024scenecraft} demonstrates conversation-driven scene revision, but a fully interactive generation loop that maintains a persistent, editable asset state remains open.

The third bottleneck---assetization---compounds the first two. Even when a method produces a visually compelling shape, the gap between its output and a deployable asset encompasses retopology, UV unwrapping, PBR separation, rigging, LOD creation, and collision fitting. Each stage has received individual attention (MeshAnything~\cite{chen2024meshanything}, Paint3D~\cite{zeng2024paint3d}, RigNet~\cite{xu2020rignet}), and recent integrated systems such as Hunyuan3D Studio~\cite{lei2025hunyuan3d} have begun to chain retopology, UV unwrapping, PBR texturing, and basic rigging within a single pipeline---yet collision mesh generation, multi-level LOD hierarchies, and physics parameter assignment remain outside the scope of any current automated system and still require manual authoring, leaving a gap that undermines the practical value proposition for professional studios. Closing this gap requires generation architectures that produce deployment-ready representations from the outset and learnable, differentiable post-processing pipelines whose losses propagate back from deployment quality metrics---import success, rendering performance, deformation fidelity---to the generator itself.

\subsection{Physically Grounded Generation}
\label{sec:challenge_physics}

Most 3D generation methods optimize exclusively for visual plausibility, producing shapes and scenes that carry no information about mass, friction, elasticity, or articulation constraints. This limits utility for any application requiring physical interaction: robotics simulation, embodied AI training, and physics-based gameplay all demand assets annotated with physical properties. PhyScene~\cite{yang2024physcene} and PhysGen3D~\cite{chen2025physgen3d} have integrated physics-aware objectives that penalize interpenetration and instability during scene synthesis, demonstrating that physical constraints can enter the generation loop rather than serving only as post-hoc validation. However, these methods address placement-level physics rather than intrinsic material properties. The broader challenge is to make physical property prediction---mass, collision shape, articulation semantics---a first-class generation output. Progress along this axis requires tighter coupling with simulation platforms such as Isaac Sim~\cite{makoviychuk2021isaacgym} and Habitat~\cite{savva2019habitat,puig2024habitat3}, which already consume structured 3D assets; the missing link is a generation pathway that produces natively simulation-compatible content.

\subsection{Toward World Models: Structured 3D as Foundation for Intelligent Simulation}
\label{sec:challenge_world_models}

Perhaps the most far-reaching open question is the role of production-ready 3D generation in constructing world models. As discussed in Section~1, video-based~\cite{bruce2024genie, valevski2025gamengen, che2024gamegenx, yu2025gamefactory} and structured~\cite{nvidia2025cosmos, yang2024unisim, deitke2022procthor, yang2024holodeck} paradigms are converging toward architectures that combine structured 3D assets for physical grounding with neural rendering for visual diversity. In this architecture, the generation methods surveyed here serve as the content supply chain: geometry synthesis, texturing, rigging, and scene composition directly determine how rich and physically faithful the resulting world model can be. Conversely, world model objectives---long-horizon consistency, interactive control fidelity, physical plausibility---provide supervision signals that can guide generation toward outputs better suited for intelligent simulation.

The connection to embodied AI follows the same logic. Training agents in platforms such as Habitat~3.0~\cite{puig2024habitat3} and Isaac Sim~\cite{makoviychuk2021isaacgym} requires large libraries of diverse, physically accurate environments, yet existing scene libraries are small, repetitive, and costly to expand manually. Scalable 3D generation offers a path to orders-of-magnitude increases in training environment diversity, positioning 3D generation as a core competence connecting vision, language, and action within a unified spatial framework.

\subsection{Scaling, Efficiency, and Societal Considerations}
\label{sec:challenge_scaling_ethics}

Scaling from single-object generation to world-level content creation introduces challenges that per-asset speed improvements alone cannot resolve. World-scale generation requires coherent composition across thousands of objects, consistent level-of-detail management, and streaming generation that produces content incrementally as users explore. Hierarchical strategies---coarse layout first, then objects, then fine detail---align naturally with the multi-scale structure of interactive worlds and offer a principled path to tractable complexity.

The increasing capability of 3D generation also raises ethical and legal questions. Training datasets aggregate assets under heterogeneous licenses, and ownership of generated outputs remains legally uncertain; photorealistic digital humans amplify deepfake risks. Responsible deployment calls for provenance tracking, watermarking, and consent-based access controls, paralleling guidelines from 2D generative AI.

A complementary tension concerns model accessibility. Leading systems increasingly migrate from open-weight releases to proprietary APIs---Hunyuan3D~\cite{zhao2025hunyuan3d} exemplifies this pattern---making independent replication and fair comparison progressively harder. Establishing open evaluation protocols and preserving publicly reproducible baselines per capability tier should be treated as a community norm.

\section{Conclusion}
\label{sec:conclusion}
This survey has organized 3D generation research around the game-asset production lifecycle, from data foundations through geometry synthesis, topology refinement, UV parameterization, PBR texturing, skeletal rigging, to scene-level composition. Three core findings emerge. First, the representational trajectory from neural radiance fields through 3D Gaussian splatting to direct mesh and structured latent generation reflects a clear convergence toward engine-native, editable outputs. Second, although production-grade solutions now exist at individual stages (learned retopology, automatic UV, PBR texturing, neural rigging), they remain isolated; the assetization bottleneck analyzed in Section~\ref{sec:challenge_pipeline} persists as the primary barrier to practical adoption. Third, the scale--quality tension in 3D datasets remains unresolved: web-scale collections lack the annotation depth (manifold topology, PBR channels, skeletal rigs) that production-ready generation demands.

Looking forward, production-ready 3D generation serves as infrastructure for intelligent simulation. World models, embodied AI, digital twins, and spatial computing all require large-scale, diverse, physically grounded 3D environments, yet the bottleneck is not rendering speed or policy learning but the scarcity of deployment-ready assets. As video-based and structured approaches to world simulation converge, the methods surveyed here will serve as the content supply chain for structured world models, while neural rendering provides visual richness that explicit representations alone cannot produce. We call on the community to establish a unified production-readiness benchmark spanning the entire pipeline, deepen collaboration between academia and the game and simulation industries, and pursue end-to-end generation systems that accept high-level specifications and output fully deployable assets---complete with topology, UV, PBR textures, rig, physics annotations, and LOD chain---as the defining challenge for the next phase of 3D generation research.

\clearpage

{
  \bibliographystyle{plainnat}
  \bibliography{references}
}

\end{document}